\documentclass[10pt,journal,compsoc]{IEEEtran}
\usepackage[english]{babel}
\usepackage{times}
\usepackage{blindtext}
\usepackage[position=top]{subcaption}
\usepackage{tabularx}
\usepackage{xcolor}
\usepackage{textcomp}
\usepackage{url}
\usepackage{multirow,booktabs}
\usepackage[skip=1mm]{caption}
\usepackage{todonotes}
\usepackage{color,soul}
\usepackage{todonotes}
\usepackage{graphicx}
\usepackage{tikz}

\usepackage{lmodern}
\usepackage{amsmath}
\usepackage{algorithm}
\usepackage{algorithmicx}
\usepackage{algpseudocode}
\usepackage[super]{nth}
\usepackage{multirow}
\usepackage{newunicodechar}
\usepackage{amsthm}
\usepackage[utf8]{inputenc}

\usepackage{array}
\usepackage{enumitem}
\usepackage{ragged2e} 

\newtheorem{thm}{Theorem}[section]
\newtheorem{lem}[thm]{Lemma}

\newcommand{\PreserveBackslash}[1]{\let\temp=\\#1\let\\=\temp}
\newcolumntype{C}[1]{>{\PreserveBackslash\centering}p{#1}}
\newcolumntype{R}[1]{>{\PreserveBackslash\raggedleft}p{#1}}
\newcolumntype{L}[1]{>{\PreserveBackslash\raggedright}p{#1}}

\newcommand{\sys}{Shufflecast~}

\begin{document}


\title{\huge Shufflecast: An Optical, Data-rate Agnostic and Low-Power Multicast Architecture for Next-Generation Compute Clusters}

\author{Sushovan Das, Afsaneh Rahbar, Xinyu Crystal Wu, Zhuang Wang, Weitao Wang, Ang Chen, T. S. Eugene Ng\\
Rice University}

\IEEEtitleabstractindextext
{\justify
\begin{abstract}

An optical circuit-switched network core has the potential to overcome the inherent challenges of a conventional electrical packet-switched core of today’s compute clusters.
As optical circuit switches (OCS) directly handle the photon beams without any optical-electrical-optical (O/E/O) conversion and packet processing, OCS-based network cores have the following desirable properties: a) agnostic to data-rate, b) negligible/zero power consumption, c) no need of transceivers, d) negligible forwarding latency, and e) no need for frequent upgrade. 
Unfortunately, OCS can only provide point-to-point (unicast) circuits. They do not have built-in support for one-to-many (multicast) communication, yet multicast is fundamental to a plethora of data-intensive applications running on compute clusters nowadays. 
In this paper, we propose Shufflecast, a novel optical network architecture for next-generation compute clusters that can support high-performance multicast satisfying all the properties of an OCS-based network core.  
Shufflecast leverages small fanout, inexpensive, passive optical splitters to connect the Top-of-rack (ToR) switch ports, ensuring data-rate agnostic, low-power, physical-layer multicast. 
We thoroughly analyze Shufflecast's highly scalable data plane, light-weight control plane, and graceful failure handling. Further, we implement a complete prototype of Shufflecast in our testbed and  extensively evaluate the network.  
Shufflecast is more power-efficient than the state-of-the-art multicast mechanisms. Also, Shufflecast is more cost-efficient than a conventional packet-switched network. 
By adding Shufflecast alongside an OCS-based unicast network, an all-optical network core with the aforementioned desirable properties supporting both unicast and multicast can be realized.

\end{abstract}
}

\maketitle
\IEEEdisplaynontitleabstractindextext

\vspace{-4mm}
\section{Introduction}

Traditional packet-switched network cores in today's compute clusters are not sustainable in the long run as CMOS-based electrical packet switches face the challenge posed by the end of Moore's Law \cite{singh2015jupiter, ballani2020sirius}. 
The power consumption of the commodity Ethernet switches escalates at a faster rate compared to the switching capacity, 
thus hindering the free scaling for next-generation compute clusters. 
For example, a $400$ Gbps Ethernet switch with Broadcom Tomahawk III chip and bare metal hardware has $10.8\times$ more power consumption per port than a $25$ Gbps Ethernet switch with Broadcom Trident III chip and similar features. 
 Optical circuit switching technologies seem to be the most promising alternative. The major advantages of such optical circuit-switched network cores over the electrical packet-switched counterparts are as follows: a) optical circuit switches (OCS) are agnostic to data-rate as they forward the incoming photons directly, b) OCS have negligible/zero power consumption because they are bufferless and their operating principles are simple (e.g., mirror rotation, diffraction etc.), c) there is no need for transceivers at the network core because of no optical-electrical-optical (O/E/O) conversion, d) OCS have negligible forwarding latency as they do not need packet-by-packet processing, 
and e) the network core does not need frequent upgrade because OCS are data-rate agnostic. 
As a result, designing next-generation compute cluster architectures with optical circuit-switched cores has been gaining significant momentum during recent years. 
 Different proposals have leveraged a wide range of OCS technologies e.g., 3D/2D MEMS \cite{mellette2017rotornet,shrivastav2019shoal,legtchenko2016xfabric,porter2013integrating},  arrayed waveguide grating
router (AWGR) \cite{ballani2020sirius,ye2016modular,xu2018podca,wu2015scalable}, free-space optics mirror assembly \cite{ghobadi2016projector} etc.

However, unlike the packet-switched network cores that can natively support one-to-many (multicast) communication, OCS-based network cores cannot inherently multicast packets to multiple destinations.
The fundamental reason is that OCS are only capable of providing point-to-point (unicast) circuit connections between source-destination pairs with some form of dynamic reconfigurability. 
Having no support for multicast is a serious technological gap, as data-intensive applications are on the rise in large-scale compute clusters and they heavily rely on iterative big-data multicasts. For instance, consider distributed machine learning (ML) workloads in compute clusters today. 
Take the LDA algorithm~\cite{blei2003latent} as an
example. Gigabytes of data representing the word distribution of all
the sampled topics are multicasted in each algorithm iteration. 
Since an LDA job runs for thousands of iterations,
multicast traffic volume can easily reach terabytes. 
Other ML examples include the Logistic Regression algorithm for Twitter spam filtering and the Alternating Least Squares algorithm for Netflix movie rating prediction~\cite{chowdhury2011managing}. Both jobs take hundreds of
iterations, and multicast communications account for 30\% and 45\% of
the job completion time, respectively. Next, consider high performance computing (HPC) workloads which include various scientific data analysis jobs~\cite{wozniak2016gimmik,itoh1995order,furrer2006covariance}. Those applications perform iterative multicasts using MPI\textunderscore Bcast \cite{mpitutorial}, 
which is a primitive in the MPI framework for one-to-many message passing. 
Consider also data mining workloads (e.g., Apache
Hive~\cite{hive-vldb}, Spark SQL~\cite{chiba2016workload}). In such
workloads, one of the most critical and time-consuming operations is
the distributed database join, in which one of the input tables is multicasted to all workers. These tables are up to $6.2$~GB in a
popular database benchmark~\cite{TPCH}.

Hence we believe, enabling high-performance multicast for next-generation compute clusters while preserving all the properties of OCS-based network core is the most necessary next step, as it will provide a crucial missing piece of the all-optical circuit-switched network puzzle. However, conventional solutions are not enough.
On one hand, application-level peer-to-peer overlays on OCS-based cores would be a zero capital-cost solution, but it would suffer from poor multicast performance and high power consumption due to redundant data transmission.
On the other hand, network-level multicast (a.k.a. IP-multicast) on a separate packet-switched core (complementing the OCS-based unicast-capable core), despite achieving ideal multicast performance, won't satisfy any of the OCS properties.

Passive optical splitter is a potentially adoptable technology which supports data-rate agnostic physical-layer multicast satisfying all the properties of OCS-based network core. However, designing a cluster-wide multicast capable network using optical splitters is not straightforward. A single giant splitter cannot span across all the ToRs to provide a cluster-wide multicast tree, because the insertion loss of a splitter proportionally increases with its fanout. No optical transceiver would be able to compensate such high insertion loss of that giant splitter. Also, splitter cannot make smart forwarding decisions when necessary, due to lack of software control.

We present a novel optical architecture called Shufflecast to support high performance multicast in next-generation compute clusters, which complements any unicast capable OCS-based network cores and preserves all the properties. Shufflecast has a unique optical-splitter topology which can scale to arbitrary network size even using small fanout splitters, ensuring data-rate agnostic multicast at scale. 
We show that ToR-to-ToR-level routing on Shufflecast can be static, yet such simplicity in routing still optimally exploits the topology and enables multiple one-to-all multicast to happen simultaneously at line-rate. 
Moreover, such static nature of routing eliminates the need for runtime ToR-to-ToR-level tree construction, group state exists only at the network edge; which makes its control plane light-weight.
Shufflecast is robust enough against single relay failure. We design a failure recovery algorithm which completely restores the reachability with graceful performance degradation. Finally, we develop a prototype implementation of Shufflecast and perform comprehensive testbed evaluation. We demonstrate that Shufflecast is up to $1.77\times$ more power-efficient compared to a peer-to-peer overlay on an OCS-based unicast network core. Also, Shufflecast is up to $1.85\times$ more power-efficient and $1.89\times$ more cost-efficient compared to IP-multicast on a minimal-layer packet-switched network core. 
Shufflecast ensures high physical-layer reliability and works well with existing transport layer protocols. Furthermore, we show that real-world high-throughput and low-latency applications can leverage and benefit from Shufflecast with only minor modifications.
\vspace{-6mm}
\section{Motivation}
\label{motivation}
\subsection{Advantages of OCS-based network core} The fundamental properties of OCS-based  network cores are: a) data-rate agnostic nature, b) negligible/zero power consumption, c) no need of transceivers, d) negligible forwarding latency, and e) no need for frequent upgrade.
OCS are agnostic to data-rate because they direct the incoming photon beams across predefined circuits irrespective of the modulation rate of the electronic signal. 
OCS intrinsically have negligible or zero power consumption due to their operating principles. For example, MEMS-based OCS consume very little power just to drive the DSP circuitry used for rotating the mirrors to setup the circuits among input/output ports. As another example, AWGR switches are fully passive (i.e., consumes no power) as they perform wavelength routing of the optical signals across the predefined input/output ports based on diffraction grating.
As OCS deal with photons, they do not need optical transceivers for O/E/O conversion. As a consequence, OCS do not need any electronic data processing or buffering which leads to negligible forwarding latency. 
Due to the data-rate agnostic property and absence of transceivers, the OCS-based network cores need not be replaced even as the network edge (ToRs and servers) is upgraded to higher speeds.
Finally, the combination of all these aspects results in OCS-based network cores to
be sustainable in the long run, while achieving close to non-blocking network performance for point-to-point (unicast) communication. Hence, there is a major momentum shift towards building such OCS-based cores for next-generation compute cluster architectures \cite{mellette2017rotornet,shrivastav2019shoal,
ballani2020sirius,legtchenko2016xfabric,
xu2018podca,ghobadi2016projector}.

\vspace{-4mm}
\subsection{Problem of OCS-based network core: Lack of multicast capability} Unlike the packet-switches, OCS are not capable of supporting point-to-multipoint (multicast) connectivity. However,  distributed ML/HPC/database applications are dominating workloads in today's compute clusters and such applications heavily rely on multicast. Hence, there is an urgent need for the next-generation compute clusters to support high performance multicast while preserving all the properties of OCS-based network core. 
Under these circumstances, the easiest approach would be to deploy the application-level peer-to-peer overlay on OCS-based cores. Here, the application organizes its processes into an overlay network and the peers distribute multicast messages as TCP-based unicast flows~\cite{banerjee2002scalable,castro2003splitstream,das2002swim, jannotti2000overcast,vigfusson2010dr,hosseini2007survey,castro2003evaluation}. 
Despite being a zero capital-cost solution with easy deployability, peer-to-peer overlay-based multicast suffers from bandwidth inefficiency because of significant data packet duplication at the end hosts and high control overhead. Such high data redundancy leads to non-negligible link stress (e.g., $1.9-10 \times$) which becomes worse with large multicast group size \cite{cao2013datacast,chu2002case,castro2003splitstream}. Even when very carefully optimized by experts, redundancy is still
at $39\%$ \cite{cao2013datacast}. Additionally, application layer overlays can lead to unpredictable latency fluctuation in relay server performance with large multicast group size~\cite{basin2010sources}. 
Based on our experiments, overlay multicast in state-of-the-art frameworks like MPI~\cite{mpitutorial} and Spark~\cite{chowdhury2011managing}, can be  $3-5.7\times$ slower than optimal (see section \ref{high_multi_low_cost}). Overlay-based multicast also suffers from high power consumption due to redundant data transmission. 

\begin{figure}
 \centering
  \includegraphics [width=3in]{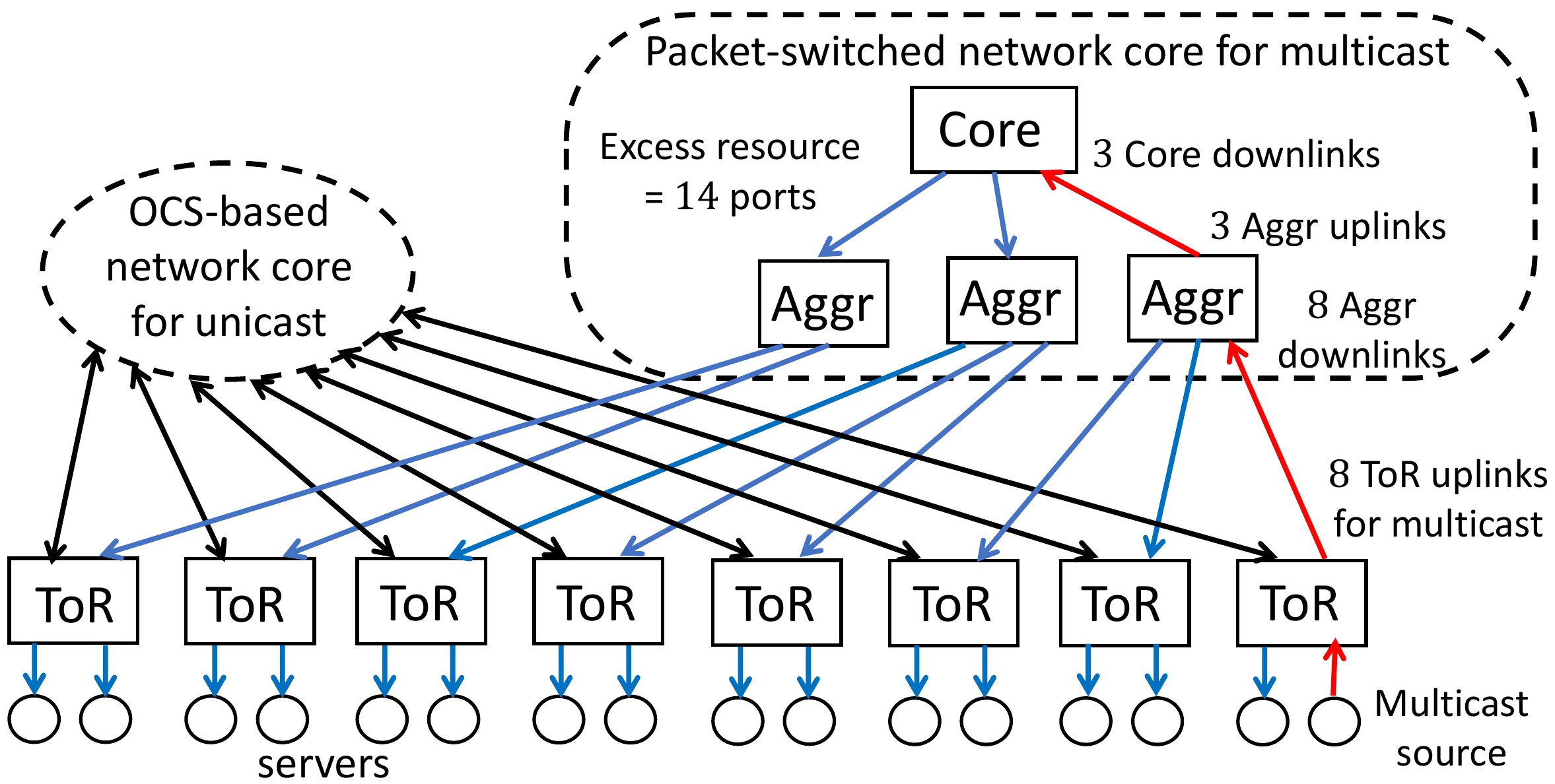}
  \caption{A ``hybrid'' network architecture: OCS-based core serves unicast and a separate packet-switched network core serves multicast traffic. 
The minimal-layer packet-switched network core requires $175\%$ excess resource to support one cluster-wide multicast tree across eight ToR ($4$-port) switches.}
 \vspace{-0.3cm}
  \label{fig:tree_multicast}
\end{figure}

Therefore, enabling high performance multicast in next-generation compute clusters while preserving all the properties of OCS-based network core is challenging. 
Conventionally we could imagine a ``hybrid'' network architecture, where OCS-based network core serves the unicast traffic and a separate hierarchical packet-switched network core serves the multicast traffic exclusively. 
Such a packet-switched network core  
would preserve the ideal multicast performance, as the packet-switches can inherently support 
IP-multicast
forwarding without any data redundancy. However, it would violate all the OCS properties, as packet switches are not agnostic to data-rate; they have high power consumption; they need transceivers, packet-by-packet processing and short-term upgrade. Moreover, such a network would have high capital cost. 
Even constructing a minimal layer packet-switched network core using identical port-count packet switches (same as ToR switches) would require non-trivial amount of electronics. To quantify such effect, we define a metric “excess resource usage” which is the ratio of extra switch ports to total ToR uplink ports, expressed in percentage. As an illustrative example, consider a simple cluster with eight $4$-port ToR switches shown in Figure \ref{fig:tree_multicast}. To support a one-to-all multicast tree using a minimal-layer packet-switched network core, we need 14 extra switch ports apart from $8$ uplink ToR ports, leading to $175\%$ excess resource usage. Similarly, a cluster with $192$ $32$-port ToR switches require at least $107\%$ excess resources to enable a one-to-all multicast. Hence, deploying such a network will not be sustainable in the long run.

\vspace{-4mm}
\subsection{Explore optical splitter technology} Fortunately there exists optical splitters, an alternative technology to enable high performance multicast without data duplication. Optical splitter is a small passive device that splits the incoming optical signal from one input fiber to multiple output fibers (defined as fanout), thus providing built-in physical layer support for line-rate multicast. Additionally, optical splitter satisfies all the properties of OCS i.e., agnostic to data rate as it has no electronic processing,  passive and no power consumption, no O/E/O conversion, bufferless and negligible latency, long term sustainable and no frequent upgrade. Furthermore, splitters are inexpensive and commercially available \cite{FS}. 

But, designing a low-diameter yet cluster-wide scalable multicast capable architecture is still an open problem, as making use of splitters have several difficulties.
 Na\"ively we could use one giant splitter to directly join all the ToRs in a cluster consuming one transceiver port from each. 
Such a design is unrealistic and practically infeasible because the insertion loss (in absolute scale) of a splitter increases proportionally with bigger fanout. Empirically, the insertion loss (in log scale) of a splitter with fanout $p$ is given by  $0.8 + 3.4 \log_2 p$ dB.  Hence, a compute cluster with 1024 ToRs would require a giant splitter of fanout 1024, having insertion loss of $34.8$ dB. Such high insertion loss cannot be compensated by any commercially available optical transceiver. A high-gain optical amplifier would be able to compensate such loss, but at the cost of higher power consumption, higher capital cost \cite{EDFA} and lower signal-to-noise ratio (SNR) at the receiver. Hence, such a network has limited scalability.  Moreover, splitter is a dumb device, i.e., it does not have the ability to make smart decisions e.g., configure the multicast trees for different sources, redirect the traffic during failure etc.
 
We design shufflecast, a highly scalable and low-diameter multicast-capable optical network architecture for next-generation compute clusters, which leverages small fanout passive optical splitters to connect the ToR ports. Thus, Shufflecast provides high performance multicast, while preserving all the OCS properties. By supporting multicast and complementing the unicast capable OCS-based network core, Shufflecast is a crucial component in the all-optical network core puzzle. In the next sections, we will show the following advantages of Shufflecast:

\begin{enumerate}[label=\alph*)]

\item Shufflecast's data plane achieves high scalablility with low network diameter (Section \ref{shuffle_data}) using small fanout splitters, ensuring data-rate agnostic multicast. As  Shufflecast can scale with limited number of ToR ports, it has low capital cost (Section \ref{high_multi_low_cost})

\item The optimal ToR-to-ToR-level routing over Shufflecast (Section \ref{shuffle_data}) supports simultaneous one-to-all multicasts at line-rate. Also, Shufflecast is power efficient compared to the conventional multicast solutions (Section \ref{high_multi_low_cost}). 

\item Due to the static nature of the routing, ToR-to-ToR-level multicast tree construction at runtime is not necessary; group state exists only at the network edge. 
Hence, the control plane of Shufflecast is very simple and light-weight (Section \ref{control}).

\item Shufflecast provides good failure resilience and graceful performance degradation after failure recovery (Section \ref{failure} and \ref{fail-eval}).  

\item Shufflecast can reliably support multiple multicast groups using existing multicast transport protocols (Section \ref{fast_control_multi_grp}). Furthermore, real-world applications can benefit from Shufflecast with minor modifications (Section \ref{high_thput}).  

  \end{enumerate}
\vspace{-3mm}
\section{Shufflecast Architecture}
\label{design}

In this section, we  discuss the Shufflecast architecture in detail with data plane design,  control plane design and failure handling.  
\vspace{-3mm}
\subsection{Data plane}
\label{shuffle_data}
In the Shufflecast data plane, passive optical splitters provide direct ToR-to-ToR connectivity.
The optical transceivers and splitters 
are co-located at the ToRs without consuming extra rack space. 

\vspace{-2.2mm}
\subsubsection{Topology}
\label{Shuffle_topology}
\begin{figure}
\centering
  \includegraphics[width=3in]{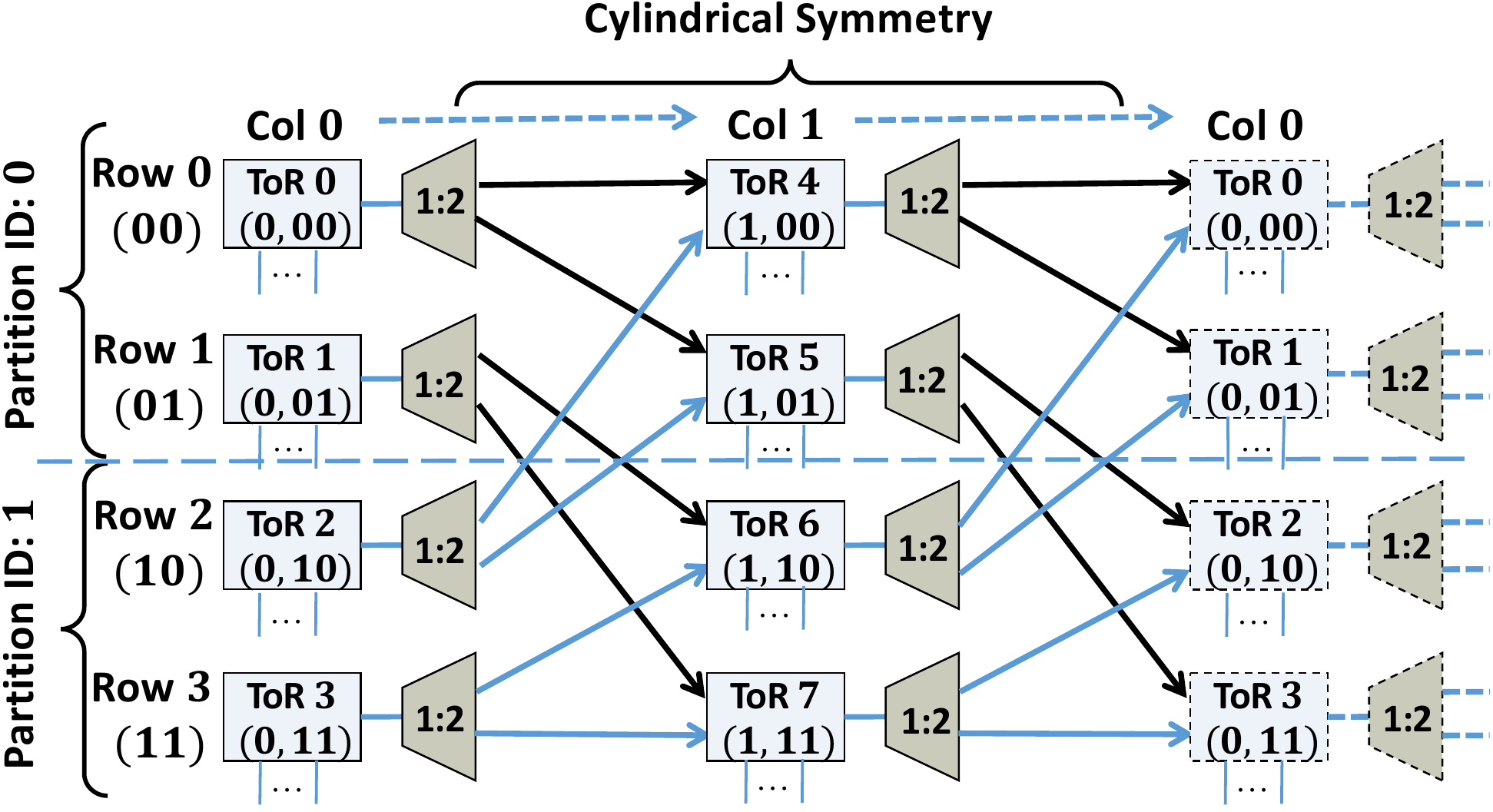}
  \caption{Connectivity of $2,2$-Shufflecast.}
  \vspace{-2mm}
  \label{fig:connectivity}
\end{figure}

The Shufflecast topology is parameterized by $p$ and $k$, where $p$ denotes the number of ToRs that a single ToR connects to via a splitter, and
$k$ is the number of {\em logical} ToR columns in the topology. In general, a $p,k$-Shufflecast has $N=k\cdot p^k$ ToR switches forming a $p$-regular graph, with each column having $p^k$ ToRs.
Figure \ref{fig:connectivity} shows an example of $2,2$-Shufflecast,
where there are 
$8$ ToRs arranged in $2$ columns, with 
$4$ ToRs per column and each ToR equipped with $1:2$ optical splitter (nodal degree $2$). More examples are in Appendix~\ref{$(2,3)$-Shuffle}.

\setlength\belowcaptionskip{-3ex}
\setlength{\belowcaptionskip}{-20pt}

\textbf{Logical ToR ID:}\label{connectivity}
We realize the Shufflecast topology  using IP-based L$2$/L$3$ Ethernet switches. The ``logical'' ToR IDs are defined to 
explain the properties of the topology and the routing scheme.
In a $p,k$-Shufflecast, the columns $(c)$ are numbered as $0,1...({k-1})$ from left to right, and the rows $(r)$ are numbered as $0,1...({p^{k}-1})$ from top to bottom. 
Any ToR 
with a decimal representation `$i$' ($i\in\left[0,{N-1}\right]$) is uniquely identified by the pair $\left(c^i,r^i\right)$ where column ID ($c^i$) is $\lfloor \frac{i}{p^k} \rfloor$ and row ID ($r^i$) denotes the $k$-tuple $p$-ary representation of $(i\bmod p^k)$ given by $[r^i_{k-1}r^i_{k-2}...r^i_1r^i_0]$. For $2,2$-Shufflecast shown in Figure \ref{fig:connectivity}, 
each ToR 
has a binary $2$-digit row ID $r_1r_0$. Considering any ToR switch e.g., ToR $6$, its column ID is $\lfloor \frac{6}{2^2} \rfloor = 1$  and row ID is the binary representation of $(6\bmod 2^2)=2$, i.e., $10$, resulting in a combined ID $\left(1,10\right)$. 

\textbf{ToR connectivity:}\label{neighbour}
We can further define the ToR connectivity pattern of Shufflecast topology  using such logical IDs. Any ToR $(c^i ,r_{k-1}^ir_{k-2}^i...r_1^ir_0^i)$ is connected to $p$ other ToRs of the next column ($c^j=\left( c^i+1 \right)\bmod k$), having the row IDs as 1 place left-shift of its own row-ID digits 
with the least significant digit $m\in \left[ 0,{p-1} \right]$(i.e., $r^j = [r_{k-2}^ir_{k-3}^i...r_0^{i}m]$).

\textbf{Partition:}\label{partition}
We {\em logically} partition the columns into $p$ regions based on the logical ToR IDs. The partition ID of each ToR is defined by the most significant digit of the ToR's $p$-ary row ID (i.e., $r_{k-1}\in\left[0,{p-1}\right]$). For the $2,2$-Shufflecast in Figure \ref{fig:connectivity}, every column has two partitions with partition IDs $0$ (ToRs $\{0,1\}$ and $\{4,5\}$) and $1$ (ToRs $\{2,3\}$ and $\{6,7\}$). All the outgoing links from partition ID $0$ are marked with darker arrows and those from partition ID $1$ are marked with lighter arrows.
The notion of partition has two important properties. a) 
 A {\em logical} partition refers to an independent resource unit (i.e., subset of relays) of Shufflecast topology, which is evident from the connectivity structure. In general, a partition containing $p^{k-1}$ ToRs is sufficient to forward the multicast traffic to all the $p^k$ ToRs of the next column.  
 b) The number of partitions in a given column dictates the degree of parallelism for Shufflecast topology. Because, the relays from different partitions of a given column can forward the multicast traffic in parallel without any interference. 
 In section \ref{rout_algo} we discuss the ToR-to-ToR-level routing scheme, which cleverly exploits such parallelism of Shufflecast topology to support multiple one-to-all multicasts simultaneously at line-rate.
 
\subsubsection{Topological properties}
\label{topo_prop}
The unique topology of Shufflecast has
some highly desirable properties such as high scalability and bounded latency. 

\textbf{Scalability and port counts:}
Shufflecast topology can scale to an arbitrary network size ($N=kp^k$) with small splitter fan-out ($p$), 
by increasing the parameter $k$ (independent of power-splitting limitations). The number of columns scales linearly ($k$) and the number of rows scales exponentially ($p^k$).
At first glance, each ToR needs $1$ transmit and $p$ receive ports. However, one transmit and one receive port can be simultaneously handled by one transceiver in practice, which leads to $p$ transceiver ports 
consumed per ToR. 
For example, a $2,2$-Shufflecast can accommodate $8$ ToRs. Similarly, a $2,3$-Shufflecast (Figure \ref{fig:connectivity_big} in \ref{$(2,3)$-Shuffle}) scales to $24$ ToRs. Both these instances only require $2$ transceiver ports per ToR.

\textbf{Hop counts:} \label{hop_count}
Leveraging the topological properties of Shufflecast, routing can be performed with low worst-case hop count ($\propto \log_p N \approx k$). 

\vspace{-1mm}
\begin{lem}
For a $p,k$-\sys all the ToRs are reachable from a given source by at most $2k-1$ hops. 
\label{hop}
\end{lem}
\vspace{-2mm}

Intuitively, we generate the multicast tree along the splitter-based connectivity from any given source ToR, and all other ToRs can be reached from the source column within two complete traversals. For example, in $2,2$-Shufflecast of figure \ref{fig:connectivity}, multicast packets from ToR $0$ can reach ToR $4$ and $5$ in $1^{st}$ hop. 
At $2^{nd}$ hop, ToR $4$ relays these packets to ToR $1$, and ToR $5$ relays to ToRs $2$ and $3$. During the second traversal, either of ToR $1$ or $3$ can relay the packets to ToRs $6$ and $7$ in $3^{rd}$ hop. Therefore, the maximum hop count is $3$. A proof is given in Appendix~\ref{lemma_proof_appen}.

\subsubsection{Multicast-aware routing} 
\label{rout_algo}

To multicast packets from a source, every ToR along the path needs to know whether packets should be relayed via its optical splitter. Our multicast-aware routing provides static ToR-to-ToR-level
relaying rules that depend only on the source ToR ID, without needing runtime switch reconfigurations.
Separately, ToR-to-server forwarding is dynamically configured based on the multicast group as needed by the applications. 

\setlength\belowcaptionskip{-3ex}
\begin{algorithm}[t]
\caption{Next-hop Relay Computation Algorithm}\label{alg:bcast_route}
\begin{algorithmic}[1]
\State $src$ = $c^s$, $r_{k-1}^s...r_{1}^sr_{0}^s$, $dest$ = $c^d$, $r_{k-1}^d...r_{1}^dr_{0}^d$
\State $cur$ = $c'$, $r'_{k-1}...r'_1r'_0$ 
\If {$c^d == c'$}
	\State $X=k$
\Else
	\State $X= \left(k+c^d-c'\right) \bmod k$
\EndIf
\If {($X==k$) OR ($r_{k-1}^d...r_X^d == r'_{k-X-1}...r'_0$)} \label{cond0}
	\State $next$ = $ \left(c'+1\right) \bmod k$, $r'_{k-2}...r'_0r_{X-1}^d$ \label{op0}
\Else
    \State $X'= \left(k+c'-c^s\right) \bmod k$
	\State $next$ = $ \left(c'+1\right) \bmod k$, $r'_{k-2}...r'_0r_{k-X'-1}^s$ \label{op1}
\EndIf
\end{algorithmic}
\end{algorithm}

\begin{figure}[t]
\centering
  \includegraphics[width=\columnwidth]{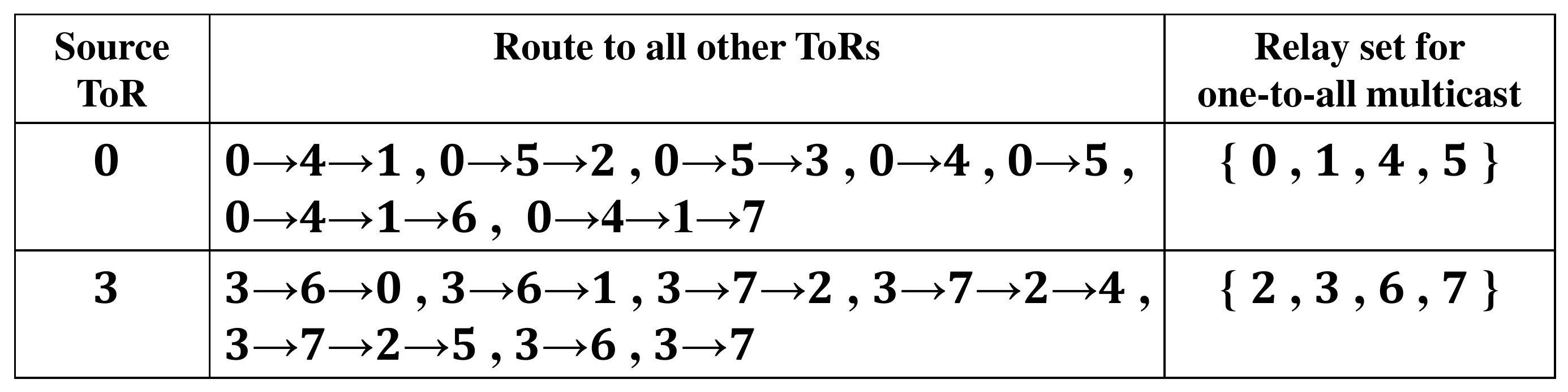}
  \caption{Relay sets for ToR-$0$ and ToR-$3$ in $2,2$-Shufflecast.}
  \label{fig:new_routing}
\end{figure}

The objective of the multicast-aware routing is to maximize the utilization of disjoint one-to-all multicast trees exploiting the degree of parallelism of the Shufflecast topology. Algorithm \ref{alg:bcast_route} illustrates next-hop relay computation. It takes the source ($src$), destination ($dest$) and current ($cur$: initialized to $src$) ToR IDs as input (lines $1$ and $2$), and 
computes the next-hop ($next$) ToR ID which acts as the relay for routing packets from that source towards the given destination. At a high level, the algorithm determines whether the destination ToR is reachable from the source ToR during the first traversal or second traversal cycle. Accordingly, it finds the next-hop ($next$) ToR ID by shifting the current ToR's row-ID to the left by one digit; and putting pre-calculated row-ID digit from either destination or source ToR ID as a least significant digit (lines $9$ and $12$). As shown in Figure \ref{fig:new_routing}, we calculate all routes and relay sets for multicast sources ToR-$0$ $\left(0,00\right)$ and ToR-$3$ $\left(0,11\right)$ of $2,2$-Shufflecast using Algorithm \ref{alg:bcast_route}.

The routing algorithm enables any source ToR to perform one-to-all multicast while choosing the relays from each column in a compact manner. More specifically, a given source ToR uses the subset of relay ToRs from each column which belong to the partition IDs defined by the source row-ID digits, termed as \textit{partition criteria}. Such selective inclusion of relays ensures the maximal utilization of the Shufflecast topology, which we generalize in the next section. As shown in Figure \ref{fig:new_routing}, ToR $0 (0,00)$ and ToR $3 (0,11)$ in $2,2$-Shufflecast relay through partition IDs $0$ (ToRs ${0,1}$ and ${4,5}$) and  partition IDs $1$ (ToRs ${2,3}$ and ${6,7}$) of both the columns respectively, maintaining the partition criteria. As a consequence, ToRs $0$ and $3$ have disjoint relay sets and they can perform one-to-all multicasts simultaneously at line-rate.

\subsubsection{Routing properties} 
\label{rout_prop} 

Shufflecast has the ability to exploit all degrees of network parallelism, with careful choices of relay ToRs, enabling  high
multicast performance. Next, we formally state the properties of multicast-aware routing with high level insights. All the proofs are in Appendix~\ref{lemma_proof_appen}.
\vspace{-1mm}
\begin{lem}
Using the multicast-aware routing for a $p,k$-Shufflecast, any given source ToR can perform one-to-all multicast following the partition criteria i.e., using the relays from each column belonging to the partition IDs predefined by its $k$ row-ID  digits. 
\label{p_criteria}
\end{lem}
\vspace{-1mm}

The intuition is from the construction of next-hop relay computation algorithm. For computing the next-hop relay ToR ID, the algorithm \ref{alg:bcast_route} carefully uses pre-calculated  source or destination ToR row-ID digits. Eventually, those source ToR row-ID digits govern the partition for choosing the relays. 

\vspace{-1mm}
\begin{lem}
Using the multicast-aware routing for a $p,k$-Shufflecast, $p$ ToRs in one column can perform one-to-all multicasts simultaneously at line-rate, $2p$ ToRs at half of line-rate, $3p$ ToRs at one-third of line-rate, and all $p^k$ ToRs in one column at $p^{k-1}$ fraction of line-rate.  
\label{simul_bcast} 
\end{lem} 
\vspace{-1mm}
The result is directly obtainable from Lemma \ref{p_criteria} and the definition of partition (Section \ref{partition}). 
Multicast-aware routing effectively exploits all degrees of network parallelism. 

\vspace{-1mm}
\begin{lem}
\label{optimality}
Multicast-aware routing is optimal in terms of minimizing the  relay usage and maximizing the number of one-to-all simultaneous multicast at line-rate.   
\end{lem}
\vspace{-1mm}

The first part of this lemma is directly obtainable from Lemma \ref{p_criteria} and properties of partition discussed in Section \ref{partition}. Any given source ToR uses one partition of relays from each column by multicast-aware routing, which indeed is the minimum number of ToRs required to reach all the ToRs in the next column. Further, the second part of this lemma is obtainable by extending this intuition along with Lemma \ref{simul_bcast}.

\vspace{-3mm}
\subsection{Control plane}
\label{control}

We assume that ToR switches support direct control of forwarding rules
(e.g., OpenFlow or P4 switches). These switches identify and forward the multicast packets sent by applications (IP datagrams with Class D destination addresses). 

\subsubsection{Static ToR-to-ToR relaying} 

For a given instance of Shufflecast, we need to apply the relay computation algorithm for each multicast source ToR once to obtain the list of relays on the
routes towards all destination ToRs. Then we 
insert one forwarding rule on these relay switches in regard to that
source ToR.
With these relay forwarding rules, data can flow from a source to
all other ToRs through the designated relays. As the forwarding rules
can be precomputed, they can be pre-installed on the ToR
switches which eliminates the need for computing routes at runtime.
Moreover, the number of such fixed rules are 
not significant compared to the memory capacity of modern switches. As discussed in Section \ref{Shuffle_topology},
each ToR in a $p,k$-Shufflecast needs to install $kp^{k-1}$ fixed forwarding rules as it relays multicast packets for $kp^{k-1}$ source ToRs. For example, a $4,4$-Shufflecast covering $1024$ ToRs needs only $256$ static forwarding rules to install on each ToR where the modern OpenFlow-based SDN switches can accommodate more than $10$k rules. Hence, the scheme is highly scalable.

\subsubsection{Application-directed ToR-to-server forwarding}

We enable dynamic ToR-to-server forwarding rule update based on application defined multicast server group membership. All the ToRs are managed by a logically centralized controller. The application interacts with the switches via the controller. When the application starts, one of its processes proactively sends the multicast group membership configuration request to the controller and waits for its response. Then the controller identifies the active servers (of that multicast group) under each ToR switch, converts them into corresponding multicast rules (capable of forwarding incoming packets to multiple ports simultaneously) and install those rules on the switches. Finally the application proceeds after getting the acknowledgement from the controller. By doing so, multicast data is confined to only the servers who belong to the respective multicast group defined by the application, which avoids unnecessary contention.

  \vspace{-3mm}
\subsection{Failure handling}
\label{failure}

Fault tolerance is another important consideration for architecture design. Next, we discuss data and control plane failure handling of Shufflecast in detail. 

\vspace{-2mm}
\subsubsection{Data plane failure handling} 

The primary sources of the Shufflecast data plane failure are bad optical transceiver, bent fiber, damaged splitter and dirty connector~\cite{zhuo2017understanding}. 
We consider any such
component failure 
as a complete failure of the associated relay. 
We discuss the performance impact of single relay failure and our re-routing algorithm to get around such a failure, as correlated multiple relay failures would be relatively rare. 

\begin{figure}
 \centering
  \includegraphics[width=0.99\columnwidth]{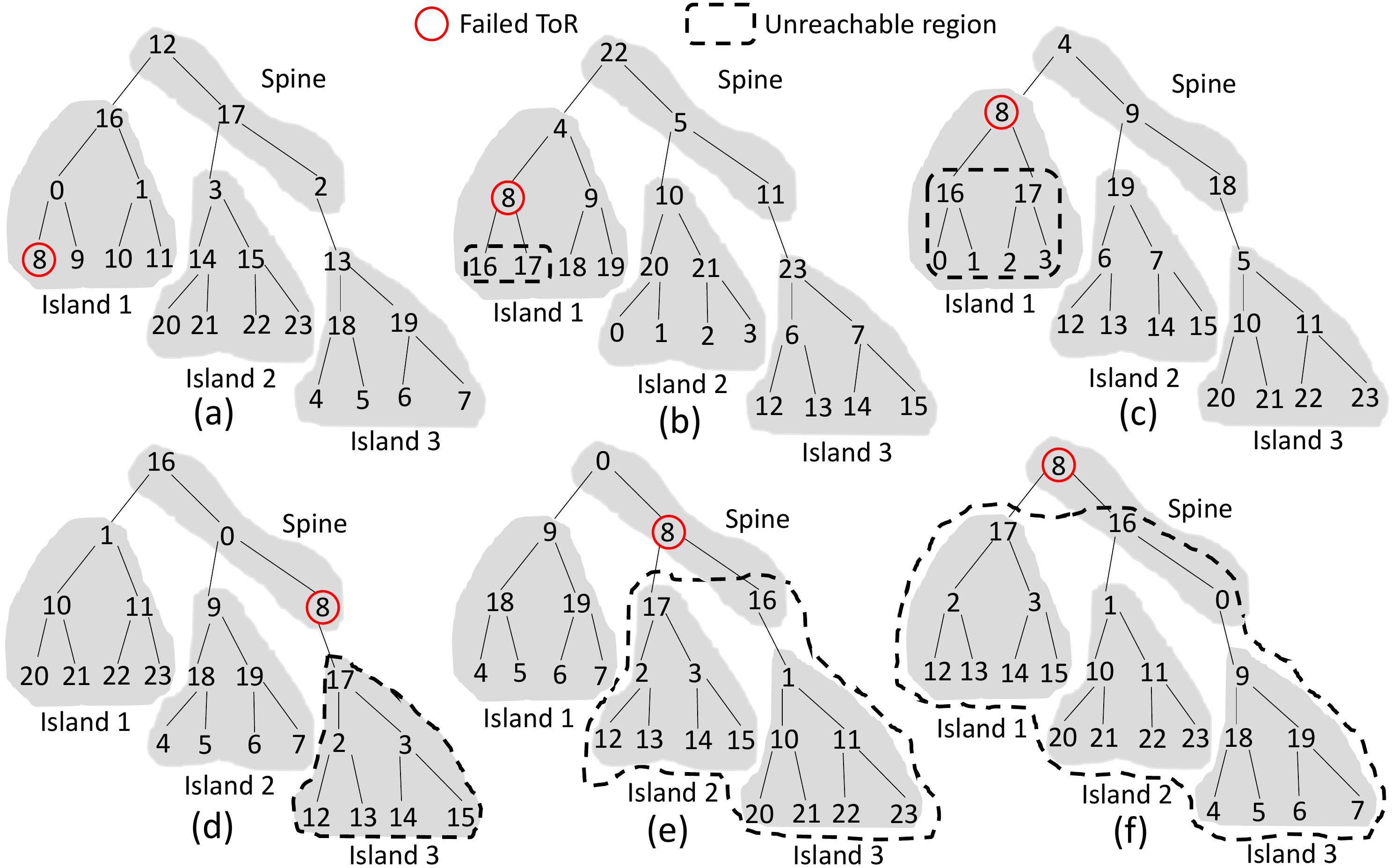}
  \caption{Different reachability scenarios when relay $8$ fails in a $2,3$-Shufflecast.  configurations (a)-(f) illustrates the unique locations of the failed relay (i.e. ToR $8$) on one-to-all multicast trees considering different source ToRs.}
  \label{fig:fail_tree}
\end{figure}

\begin{figure}
\centering
  \includegraphics[width=3in]{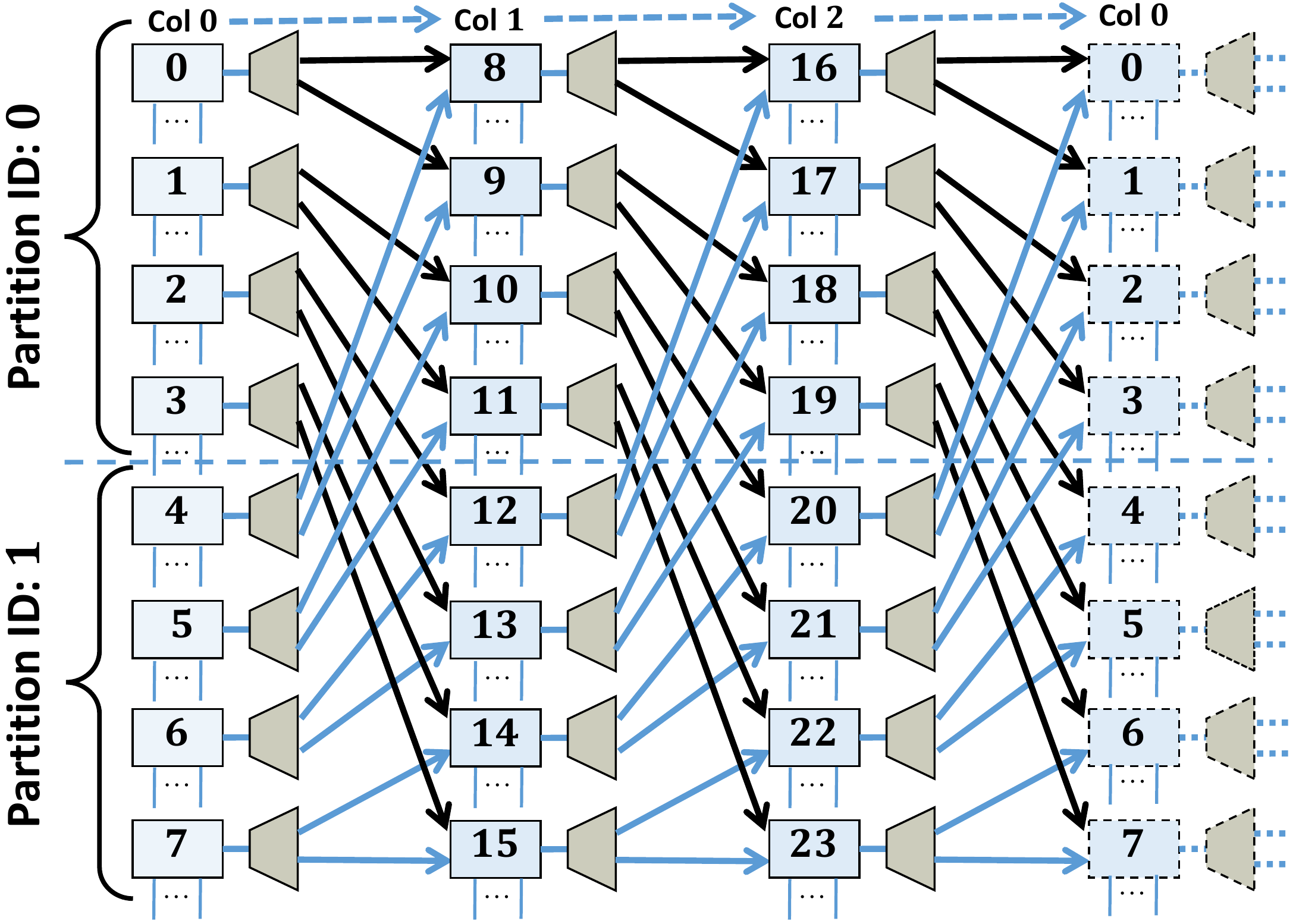}
  \caption{Connectivity of $2,3$-Shufflecast.}
  \vspace{-1mm}
  \label{fig:connectivity_big}
\end{figure}

\textbf{Reachability impact of single relay failure:} First we model the reachability impact of single relay failure on $p,k$-Shufflecast. 
Figure \ref{fig:fail_tree} illustrates different reachability scenarios for an example case and provides the intuition to  formulate the general case. Consider when ToR relay number $8$ fails in a $2,3$-Shufflecast (Figure~\ref{fig:connectivity_big} in \ref{$(2,3)$-Shuffle}). As shown in Figure \ref{fig:fail_tree}, there are six configurations ((a)-(f)) showing unique locations of the failed relay $8$ on one-to-all multicast trees of different source ToRs. All these multicast trees have similar structure; a major spine consisting of three (i.e., $k$) ToRs with source ToR as the root and one  perfect binary (i.e., $p=2$) subtree (defined as \textit{islands}) of height three (i.e., $k$), hanging from each ToR in the spine.
As we vary the source ToR, the location of the failed relay on the multicast tree varies ($24$ different locations for $24$ possible sources) and correspondingly that leads to one of these six configurations along with certain number of unreachable ToRs. 

Configuration (a) shows the case where the failed relay $8$ is a leaf in island $1$, i.e., ToR $8$ does not relay the multicast packet for that source and there are $12$ such leaf locations across three islands. Hence, there are $12$ source ToRs for which there will be no impact on reachability if relay $8$ fails.
In configuration (b), the failed relay $8$ is located at one-level above the leaf in island $1$, i.e., ToR $8$ relays the multicast packet to two (i.e., $p$) other non-relay ToRs (leafs). As there are $6$ such possible locations across the three islands, there exists $6$ source ToRs which can't send multicast data to $2$ leaf ToRs (marked with dashed contour) if the relay $8$ fails. Similarly in configuration (c), the failed relay $8$ is the root of island $1$. Hence the number of unreachable ToRs is $6$ (i.e., $p+p^2$) and $3$ source ToRs will have such impact, as there are $3$ such equivalent locations across the islands. 

Next, in configurations (d)-(f), the failed relay $8$ is located on the major spine of the multicast tree. As these locations are unique, there is a unique source associated with each of these cases. Specifically in configuration (d), the source ToR is $16$ and the failed relay $8$ is at the lowest level of the spine. Hence, ToR $16$ can't send multicast data to all $7$ (i.e., $p^k-1$) ToRs in island $3$. Similarly in  configuration (e), 
all the ToRs in island $2$ and $3$ along with the lowest relay of the spine (i.e., total $2p^k-1=15$) are unreachable from the source ToR $0$. Finally, configuration (f) shows the trivial case where failed relay $8$ is the source i.e., root of the multicast tree. Hence, all $kp^k-1=23$ other ToRs are unreachable from ToR $8$. 
Extending this idea, we compute the distribution of reachability impact of single relay failure on $p,k$-Shufflecast, which we further evaluate in Section \ref{fail-eval}.

\vspace{1.5mm}
\begin{algorithm}
\caption{Single Relay Failure Recovery Algorithm}\label{alg:fail_recov}
\begin{algorithmic}[1]
 \State $failed_{relay}$ = $c$, $r_{k-1}r_{k-2}...r_{1}r_{0}$ 
\State $y = \left(r_{k-1}+1\right)\bmod p$
\State $mirror_{failed}$ = $c$, $ yr_{k-2}...r_{1}r_{0}$        
\State Deactivate all relaying rules on $failed_{relay}$
\State Activate all $failed_{relay}$ rules on $mirror_{failed}$
\State $precedent_{relay}$ = $\left(c-1\right) \bmod k$, $r_{0}y...r_{2}r_{1}$ 
\State $y' = \left(r_{0}+1\right)\bmod p$
\State $mirror_{precedent}$ = $\left(c-1\right)\bmod k$, $y'y...r_{2}r_{1}$
\For{$i \gets 1$ to $\left(k-1\right)$}                    
 \State $n_i$ = $\left(c-i\right) \bmod k$, $r_{i-1}...r_0r_{k-1}...r_{i}$ 
 \algorithmiccomment{\textit{ Circular right shift of $failed_{relay}$ row ID by $i$ positions} }
 \State Deactivate relaying for $n_i$ on $precedent_{relay}$
 \State Activate relaying for $n_i$ on $mirror_{precedent}$
\EndFor
\end{algorithmic}
\end{algorithm}

\textbf{Single relay failure recovery:}
For $p,k$-Shufflecast, a given ToR in any column is directly connected from $p$ ToR relays (one from each partition) of the previous column. For example, in $2,3$-Shufflecast (Figure~\ref{fig:connectivity_big} in \ref{$(2,3)$-Shuffle}), ToR relays $0\left(0,000 \right)$ and $4\left(0,100 \right)$ are situated at \nth{0} location of partition IDs $0$ and $1$, respectively, and both are connected to ToR $8\left(1,000 \right)$. We define these ToR relays as ``mirrored relays'', where their row-ID digits are the same except the most significant digit which dictates the partition. 
Note that there exist more than one path to reach a set of ToRs from a given source, allowing Shufflecast to reroute packets upon relay failure. Algorithm \ref{alg:fail_recov} shows how to handle a single relay failure for $p,k$-Shufflecast. 
Depending on the failed relay ToR ID, we need to deactivate some relaying rules on two specific ToR relays (including the failed relay) and activate those on two other ToR relays, regardless of network size.

\begin{figure}[t]
 \centering
  \includegraphics[width=\columnwidth]{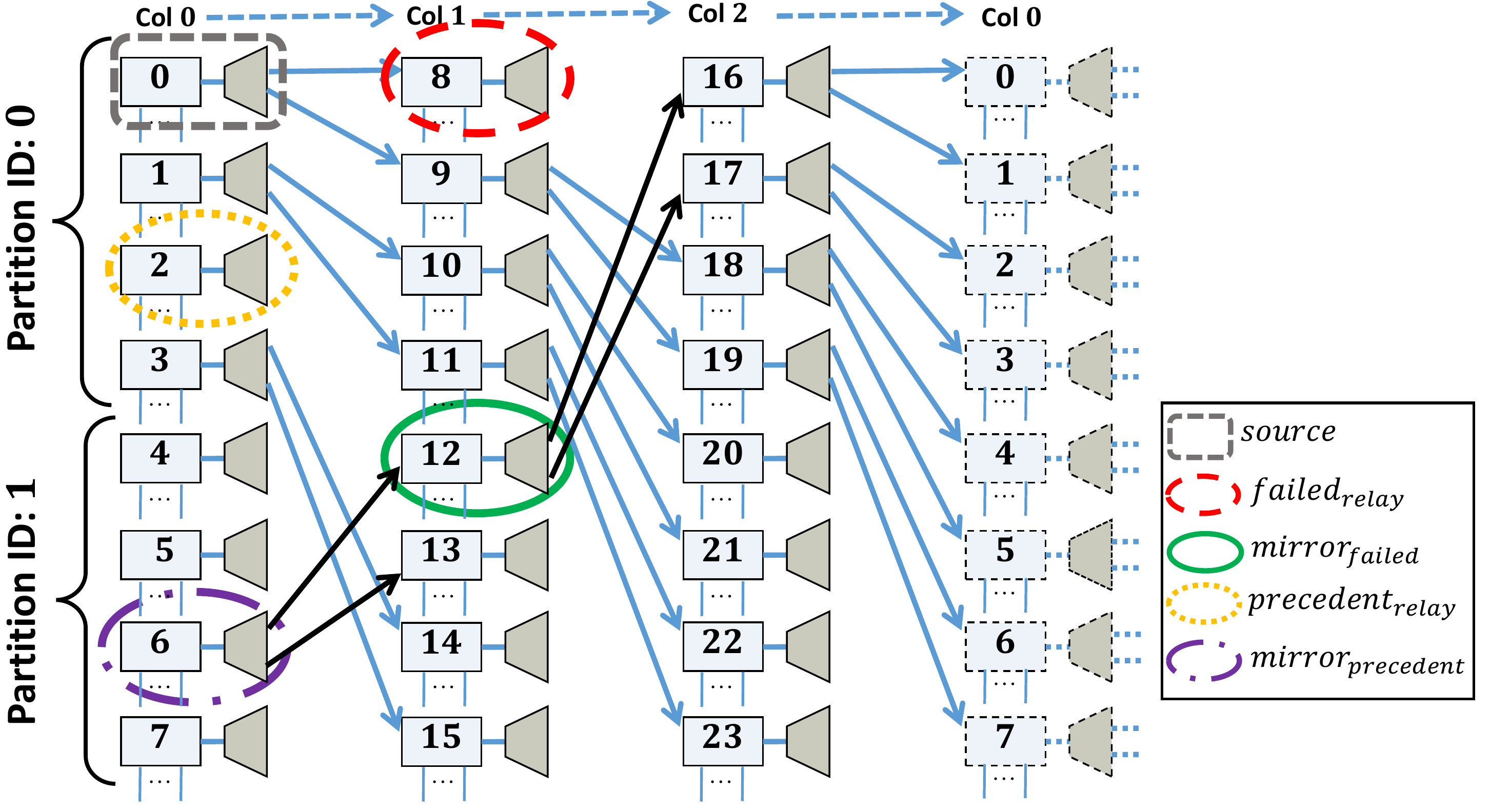}
  \caption{Failure recovery in a $2,3$-Shufflecast when ToR $0$ needs to make a one-to-all multicast while ToR $8$ fails.}
 \vspace{-0.1cm}
  \label{fig:fail_2}
\end{figure}

We explain the algorithm using the example below. Consider a $2,3$-Shufflecast, where relay $8\left(1,000\right)$ fails ($failed_{relay}$) and source $0\left(0,000\right)$ needs to perform one-to-all multicast. Based on Algorithm \ref{alg:fail_recov}, the four specific ToRs are marked in Figure \ref{fig:fail_2}, for which the relay rules will be affected. All relaying rules on failed relay $8$ are deactivated and its mirrored relay $12\left(1,100\right)$ ($mirror_{failed}$) activates those rules on its behalf (lines $1$-$5$). Additionally, the precedent relay $2\left(0,010\right)$ ($precedent_{relay}$) deactivates the relaying rules of  
a subset of source ToRs and it's mirrored relay $6\left(0,110\right)$ ($mirror_{precedent}$) activates those rules (lines $6$-$12$).

Note that, only activating the relay rules on $mirror_{failed}$ on behalf of $failed_{relay}$ is not enough. Because, after the first traversal cycle through all the columns, packets from ToR $0$ can only reach to the ToRs of partition ID $1$ (ToRs $4,5,6$ and $7$) at its own column; the ToRs from its own partition ID $0$ (i.e., ToRs $1,2$ and $3$) have not received them yet. Unfortunately, none of those relays from partition ID $1$ can forward the packets as per the routing rule. Similar situation happens for source $16\left(2,000\right)$ too. Specifically, the relay  $12$ ($mirror_{failed}$) cannot get the packets from its designated precedent relay $2$ ($precedent_{relay}$). Hence, relaying of source $0$ and $16$ ($n_1$ and $n_2$ respectively, at line $10$ inside the loop) are deactivated on relay $2$, while relay $6$ ($mirror_{precedent}$) activates those rules on its behalf. Now, ToR $0$ can successfully perform one-to-all multicast, where the outgoing links from newly activated relays are marked with darker arrows and all other required links are marked with lighter arrows.

Thus, Shufflecast can recover 100\% reachability from a relay
failure (except for the servers under the ToR of the failed relay can no longer be multicast sources) by
re-routing packets. Moreover, such failure recovery results in graceful performance degradation, evaluated in section \ref{fail-eval}.

\vspace{-2mm}
\subsubsection{Control plane failure handling}

Controller failure does not affect ToR-to-ToR forwarding in Shufflecast, as those relaying rules are static and pre-installed offline. 
However, it affects the server-level multicast group membership  configuration, as Shufflecast still needs dynamic application-directed ToR-to-server forwarding update at runtime. To handle such controller failure, the logically centralized controller can be realized as a small cluster of controllers, where one can act as primary controller and others can be as backup controllers. When the primary controller fails, a backup controller can be elected as the leader, which can be used by the application for 
runtime switch configuration.

\vspace{-2mm}
\section{Discussions}
\label{discussions}

In this section, we discuss several practical advantages in the Shufflecast architecture. 

  \vspace{-3mm}
\subsection{Leveraging idle edge bandwidth}
\label{unused_port}

Shufflecast can potentially leverage idle edge bandwidth,  
as often there exists unused switch
ports at ToRs due to design constraints on space, power, and network oversubscription.
This observation is first made by recent
works~\cite{chatzieleftheriou2018larry, CCCA, ohsita2015optical} and
confirmed by large network operators we consulted.
Additionally, we conduct an analysis to quantify the likelihood of unused ToR ports (details in Appendix~\ref{appendixunusedport}).
We consider a wide range of network configurations. The results show that unused ports, as well as a large amount of unused bandwidth, often exist. The existence of $2$+ 
unused ports and $100$ Gbps of unused bandwidth can be seen in nearly $79\%$ and $73\%$ of the cases, respectively. Under $1:1$ oversubscription, $54\%$ of cases have at least $10$ unused ports and $500$ Gbps of unused bandwidth. We also observe that the likelihood of having unused ports do not correlate with oversubscription ratios, rack sizes, and server port speeds etc., indicating that unused ports can exist throughout the continuum of configuration choices.

  \vspace{-3mm}
\subsection{Simplifying network management}
Shufflecast incurs very little need for runtime switch configurations
as it uses static optimal
ToR-level routing rules. Except for the forwarding behaviors to end
hosts at the ToRs, all ToR-to-ToR forwarding rules are precomputed and
pre-installed on switches. These
preconfigurable and static switch actions make Shufflecast
much less prone to configuration errors, which is the primary source
of network management complexities.
In addition, the physical wiring of Shufflecast is easy to deploy. For a $p,k$-Shufflecast topology, the optical transceivers and splitters are co-located at the ToRs, meaning that we only need to install $p$ incoming and outgoing optical fiber cables. In terms of wiring, the mapping from the logical ToRs to physical ToR locations is based on the logical column-wise placement, bundling fibers across partitions. Also, most physical wiring is between the adjacent physical rows of racks, and the length of fibers would not incur significant attenuation ($0.36$ dB/km at $1310$ nm \cite{FS}).

\vspace{-2mm}
 \subsection{End-to-end reliability} \label{reliability}
  
Shufflecast is dedicated to multicast traffic and leverages optical splitters to enable physical-layer multicast. Below we concretely argue how Shufflecast can ensure reliability from different aspects.

\textbf{a) Physical layer reliability:} Typically, the chances of packet loss in the optical devices are extremely rare. The optical transceivers have bit-error rate less than $10^{-12}$. Even though passive optical splitters have insertion loss, the optical link can be made completely lossless when choosing compatible optical transceivers with a feasible power budget (Table \ref{tab:cost_table_fixed}). Moreover, as shown in Section \ref{failure}, Shufflecast can gracefully handle and reroute traffic in presence of single relay failure. Hence, Shufflecast has inherent physical layer reliability.

\textbf{b) Higher layer reliability:} In presence of multiple applications, the occasional packet losses in Shufflecast links can be handled by transport layer solutions such as NORM \cite{norm}, an off-the-shelf reliable multicast protocol enabled with congestion control \cite{widmer2001extending,widmer2006tcp}. As shown in Section \ref{concur_grp}, Shufflecast can handle concurrent multicast applications using NORM with high reliability. Additionally, multiple applications can also coordinate based on the explicit knowledge of the topology, static relaying pattern and design capacity of Shufflecast network. For example, two applications can inject multicast traffic simultaneously at line-rate if they use disjoint partitions of Shufflecast; otherwise, they can take turn at line-rate based on their arrival time (FCFS) if they have common relays, thus maximizing the network utilization and minimizing packet losses between ToR-to-ToR links.

\vspace{-2mm}
\section{Implementation} 
\label{implementation}

We implement a prototype of $2,2$-Shufflecast in our testbed. Our setup uses $3$ OpenFlow switches,
$8$ optical splitters ($1:2$), and $16$ servers. We divide logically $2$ OpenFlow switches to emulate $4$ ToR switches each, 
and $2$ servers are connected to each logical ToR. We wire the Shufflecast network using optical splitters on these $8$ logical ToR switches. 
The 3rd OpenFlow switch is used for comparative evaluation, it connects to the logical ToRs, creating a $2$-layer full-bisection bandwidth network across ToR switches and emulating a non-blocking network core. Each server has $6$ 3.5GHz CPU cores with $12$ hyperthreads and $128$~GB RAM. All connections are $10$~Gbps Ethernet. 
To minimize the number of ports used, while wiring the $2,2$-Shufflecast, at each logical ToR switch we 
connect the outgoing fiber (to its own splitter) and one of the $2$ incoming fibers (from $2$ other splitters) 
to a single transceiver port.
Thus, each logical ToR consumes only $2$ transceiver ports (optimal for $2,2$-Shufflecast). The forwarding rules are installed on the switches using the Ryu OpenFlow controller~\cite{ryu}, running on one of the servers. 

The controller program consists of two parts. The first part 
runs Algorithm~\ref{alg:bcast_route} (Next-hop relay computation algorithm) and pre-installs the static ToR-to-ToR forwarding rules for $2,2$-Shufflecast ($<100$ lines of python code). The second part translates application-based multicast group membership information into the ToR-to-server multicast rules and installs them on the switches at runtime ($<30$ lines of python code). We make simple modifications to applications to interact with the controller program ($\approx 10$ lines of C++ code).
\vspace{-3mm}
\section{Evaluation} 
\label{evaluation}

In this section, we present comprehensive testbed experimental results
to demonstrate that Shufflecast can achieve a) line-rate multicast throughput with low power consumption and capital cost, b) high end-to-end reliability while supporting concurrent multicast groups, c) high robustness against single relay failure and graceful performance degradation after failure recovery and d) improved application performance for both high-bandwidth and low-latency applications. 

\vspace{-3mm}
\subsection{Shufflecast achieves line-rate multicast performance with low power consumption and capital cost}
\label{high_multi_low_cost}

\begin{figure*}[t!]
  \centering
    \includegraphics[width=6.9in]{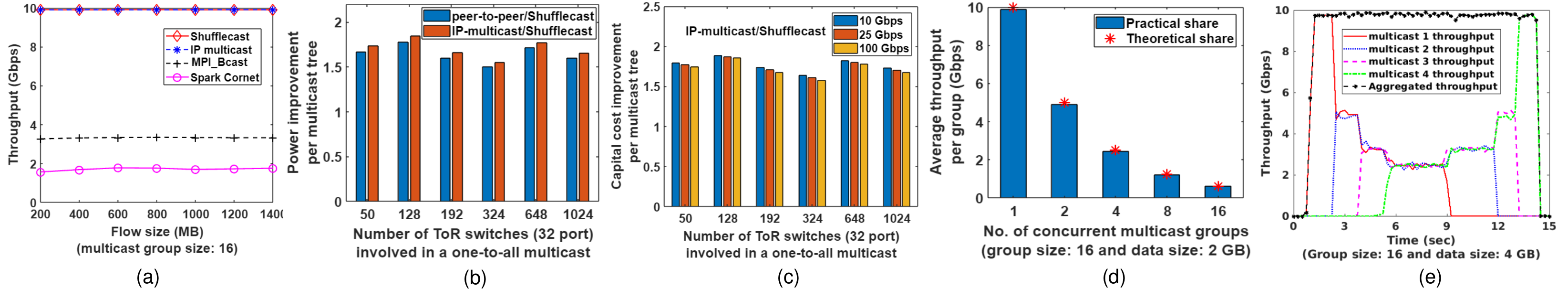}
    \caption{(a) Throughput for different multicast mechanisms (Shufflecast, IP-multicast and peer-to-peer overlay) across different data size for a $1:15$ multicast flow. Shufflecast achieves the line-rate multicast performance, (b) Improvement in power consumption per one-to-all multicast tree of Shufflecast compared to peer-to-peer overlay (on optical circuit-switched network core) and IP-multicast (on minimal layer packet-switched network core) with scale. The improvement factor is the same across different data rates ($10$ Gbps, $25$ Gbps, $100$ Gbps), as it only depends on the relative count of switch ports (same as transceivers), (c) Improvement in capital cost per one-to-all multicast tree of Shufflecast compared to IP-multicast (on minimal layer packet-switched network core) with scale at different data rates ($10$ Gbps, $25$ Gbps, $100$ Gbps), (d) Practical and theoretical average multicast throughput per group with varying number of concurrent multicast groups on Shufflecast, (e) Throughput of multicast flows launched in a staggered way on Shufflecast.}
    \label{fig:data_control}
\vspace{-2.5mm}
\end{figure*}

We perform benchmarking experiments to evaluate the multicast performance of Shufflecast. Also, our analysis show that Shufflecast is power and cost efficient across network scale.

\textbf{a) Multicast performance of Shufflecast vs. state-of-the-art multicast mechanisms:}
\label{high_multicast} For the benchmarking experiments, our baseline mechanisms are 
state-of-the-art multicast solutions i.e., 1) peer-to-peer mechanisms such as MPI\textunderscore Bcast~\cite{mpitutorial} and Spark-Cornet~\cite{chowdhury2011managing} and 2) IP-multicast. For both the baselines, we use the full-bisection bandwidth network in the testbed to measure their ideal maximal performance.

We perform a $1:15$ multicast with varying data size (from $200$ MB to $1.4$ GB) and measure the multicast reading time (i.e. the duration between receiving program issues reading request and finishes reading it). 
Figure \ref{fig:data_control}(a) shows the 
multicast throughput defined as the ratio of multicast data size to multicast reading time. We observe that Shufflecast achieves line-rate multicast throughput, same as the upper-bound performance of IP-multicast (over full-bisection bandwidth network), irrespective of the multicast group size. We also observe that, even without any competing traffic on full-bisection bandwidth network, both MPI\textunderscore Bcast and Spark-Cornet achieve the multicast throughput only upto $35\%$ and $20\%$ of the line-rate throughput across data size, which is far from optimal.

In Spark-Cornet, a node first locates a block of data it needs from
another node then performs a block transfer. We observe that although
each individual block transfer can reach near line-rate throughput,
far more time is taken up by control communications to locate and wait for data blocks, which becomes the bottleneck for overall throughput. MPI\textunderscore Bcast adopts different approaches based on multicast data size \cite{mpitutorial,bao2017icast}. For comparatively smaller data size, MPI\textunderscore Bcast uses binomial tree approach. In the first round, the multicast sender process sends data to one receiver. In the second round, these two processes send the same data to one additional receiver each and so on. For the medium and bigger data sizes, MPI\textunderscore Bcast adopts scatter + altogether approach. The  altogether is realized by recursive doubling or ring algorithm, where the data is pipelined from one node to the next. In this case, the software handling of data from input to output and the need to ensure reliability across the pipeline become the bottleneck for overall throughput.

\textbf{b) Power consumption analysis:}
Shufflecast is power efficient compared to both a) peer-to-peer overlay multicast and b) IP-multicast. Ethernet switch ports and the optical transceivers consume power. Passive splitters and fiber optic cables do not consume any power. We count the number of active switch ports and transceivers (similar methodology as \cite{ballani2020sirius}) involved in one cluster-wide multicast tree for all three network architectures. 

For peer-to-peer overlay multicast on optical circuit-switched core, we assume the lowest possible power  consumption, where the data propagates through a chain across all the ToR switches at line-rate. Thus it consumes two switch ports (with two transceivers) from each ToR (both receive and transmit). For IP-multicast, we consider the minimal-layer packet-switched network core with identical port-count packet switches as shown in Figure \ref{fig:tree_multicast}. Such a network would consume excess switch ports (with same number of excess transceivers) in addition to one port (with one transceiver) per ToR. Although for IP-multicast, the data can be instantaneously forwarded from one port to multiple ports in a switch, each port still needs to physically transmit the data to other switches. Thus, more active transmissions result in high power consumption. Finally, Shufflecast requires two active ports (with two transceivers) on each relay ToR (both receive and transmit) and one port (with one transceiver) on each non-relay ToR (only receive) to realize a one-to-all multicast tree. Shufflecast saves the number of active port (and transceiver) usage significantly, because it needs only one transmit port (with one transceiver) on any relay ToR to send the data into optical splitter. Then the splitter performs physical layer multicast without consuming power. For a simple example, in a cluster with $8$ ToRs ($4$-port switches), the number of active ports to support a one-to-all multicast (one server per ToR) will be $16$, $22$ and $12$ for peer-to-peer, IP-multicast and Shufflecast respectively. The corresponding transceiver count will also be the same. 

To evaluate the power consumption, we choose the number of ToRs in such a way that it can be realized with some instance of $p,k$-Shufflecast e.g., $50 \equiv$ $5,2$-Shufflecast, $128 \equiv$ $8,2$-Shufflecast and so on. 
The typical power consumption values \cite{FS} of different Ethernet switch ports and optical transceivers are given in Table \ref{tab:cost_table_fixed}. For Shufflecast we consider the optical transceiver having sufficient power budget to compensate the insertion loss of different optical splitters. As shown in Figure~\ref{fig:data_control}(b), Shufflecast is  $1.5-1.77\times$ more power efficient than peer-to-peer overlay. Note that for peer-to-peer overlay, we consider the active power consumption only from the network. But in reality, the power consumption will be even more because it also involves the host peers (servers) to receive and transmit the multicast data repeatedly. Also, Shufflecast is $1.55-1.85\times$ more power efficient than IP-multicast over the minimal-layer packet-switched network core. The improvement factors are the same across different data rates, as it only depends on the relative count of switch ports (same as transceivers). 

\begin{table}
\vspace{3mm}
\centering
  \includegraphics[width=3.4in]{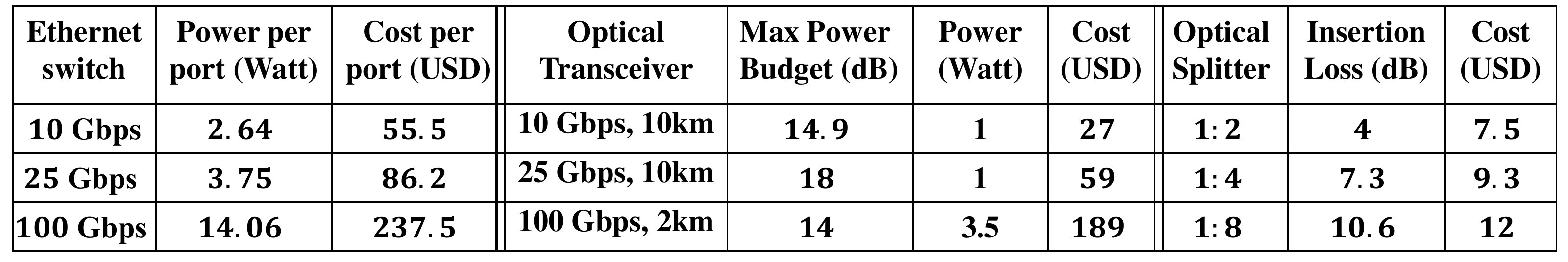}
  \vspace{3mm}
  \caption{Most recent power consumption and cost values  of different components.}
  \label{tab:cost_table_fixed}
\end{table} 

\textbf{c) Capital cost analysis:}
The deployment of Shufflecast incurs very little extra hardware cost since the optical devices including passive optical splitters, optical transceivers, and fiber-optic cables are all inexpensive. 
For a $p,k$-Shufflecast, each ToR requires one optical splitter, $p$ optical transceivers, $p$ outgoing fiber cables and $p$ switch ports. 
Table \ref{tab:cost_table_fixed} summarizes the most recent costs \cite{FS} of different components.  
The approximate cost of a duplex single-mode fiber per $100$ meter is $37.37$ USD. 
Given a $4,4$-Shufflecast (spanning $1024$ ToRs) with $100$ meter fiber optic cable as an example, the capital cost per ToR are approximately $487-1867$ USD across different data rates, which is fairly inexpensive for large clusters. 
Figure~\ref{fig:data_control}(c) shows the improvement in capital cost per one-to-all multicast tree of Shufflecast compared to IP-multicast (on minimal layer packet-switched network core) with scale at different data rates ($10$ Gbps, $25$ Gbps, $100$ Gbps). We consider the necessary components involved in one cluster-wide multicast tree for both Shufflecast (switch ports, transceivers, splitters and fiber-optic cables) and IP-multicast (switch ports, transceivers and fiber-optic cables) architectures. Based on our evaluation, Shufflecast is $1.57-1.89\times$ more cost efficient compared to IP-multicast over minimal-layer packet-switched core, across different network scale and data rates. 
We observe that the improvement factor decreases slightly with higher data rate. The reason is that switch port and transceiver costs are data-rate dependent and start dominating the fiber cost (data-rate independent) at higher data rate. As a result, the higher fiber cost for IP-multicast matters less at higher data rates. If the costs of higher speed switch port and transceiver continue to rise while fiber/splitter cost remain constant, the improvement factor will converge to the relative count of switch ports (same as transceivers) i.e., $1.55-1.85\times$.

\vspace{-3mm}
\subsection{Shufflecast achieves high reliability while supporting concurrent multicast groups with negligible overhead}
\label{fast_control_multi_grp}

We measure the responsiveness of Shufflecast control plane and experimentally demonstrate that Shufflecast achieves high reliability in presence of concurrent multicast groups using off-the-shelf transport layer solutions \cite{norm}.

\textbf{a) Shufflecast has highly responsive control plane:} Although Shufflecast has pre-installed static ToR-to-ToR relaying rules, application-directed dynamic ToR-to-server multicast forwarding rule update is required before the multicast starts (Section \ref{control}). Based on our measurement, such a multicast rule update on a Quanta T3048-LY2R OpenFlow switch only takes $0.6$ msec. Moreover, Shufflecast controller sends parallel requests to the ToRs simultaneously. For big data applications, such latency is negligible compared to their multicast durations, which can easily reach tens of seconds (Section \ref{high_thput}).

\textbf{b) Shufflecast achieves high reliability while supporting concurrent multicast groups:}\label{concur_grp} We perform multicast of $2$ GB data size over $2,2$-Shufflecast with a group size of $16$ ($1:15$ multicast) using NORM \cite{norm}, a well-known off-the-shelf reliable multicast solution. NORM \cite{norm} is a NACK-based reliable multicast protocol enabled with forward error correction (FEC) and the TCP-Friendly Multicast Congestion Control (TFMCC) scheme \cite{widmer2001extending,widmer2006tcp}. We vary the number of concurrent multicast groups from $1$ to $16$ by running parallel norm sessions on each destination server and invoking the corresponding number of servers as multicast senders. We observe that all the multicast flows get close to fair-share throughput at steady state and the packet loss is below $0.28\%$. 
Figure \ref{fig:data_control}(d)
shows that the observed average multicast throughput per group at steady state is almost same as the theoretical fair-share. Therefore, the aggregate network throughput in presence of such concurrent multicast groups is always close to line-rate.
Next, we launch $4$ multicast flows (group size is $16$ and data size is $4$ GB) with a  progressive staggering of $1.5$ sec. 
Figure \ref{fig:data_control}(e) 
shows the individual flow throughput and aggregate network utilization variation with time. We observe that multicast flows achieve their fair-share quickly and the overall network utilization is close to line-rate. 

\begin{figure*}[t!]
  \centering
    \includegraphics[width=6in]{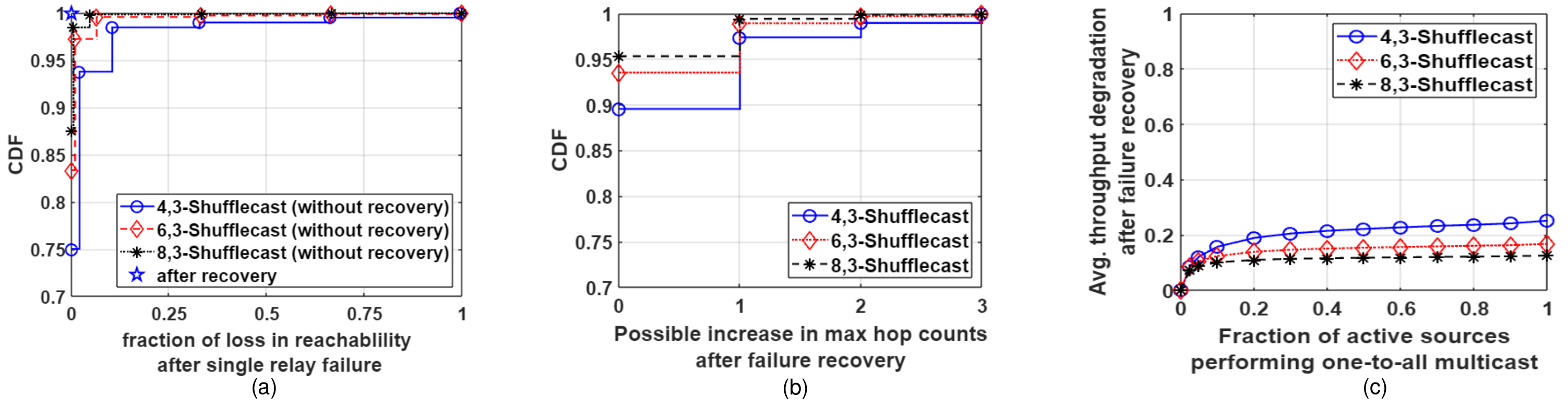}
    \caption{(a) CDF of fraction of loss in reachability of Shufflecast under single relay failure. Without failure recovery, there is no reachability impact for majority of sources performing one-to-all multicast. With failure recovery, the reachability is completely restored, (b) CDF of excess latency (in terms of max hop count) after single relay failure recovery in Shufflecast. After failure recovery, latency is unchanged for majority of the sources performing one-to-all multicast, (c) Average throughput degrades gracefully after single relay failure recovery in Shufflecast, with varying the number of active sources performing one-to-all multicast.}
    \label{fig:failure}
\vspace{-2.5mm}
\end{figure*} 

\subsection{Shufflecast achieves high robustness against single relay failure and graceful performance degradation after failure recovery}
\label{fail-eval}

We evaluate the reachability impact on Shufflecast under single relay failure. We also evaluate the impact of latency and throughput degradation of Shufflecast after enabling the single relay failure recovery.

\textbf{a) Shufflecast is robust enough against single-relay failure:} 
Based on our reachability analysis (Section \ref{failure}), we compute the distribution of reachability impact after a single relay failure on $p,k$-Shufflecast. 
Figure \ref{fig:failure}(a)  
shows the distribution for different Shufflecast instances. We observe that, the majority of sources does not have any impact in reachability under a single relay failure even before enabling the failure recovery. Also, the size of this majority increases with bigger network scale. As shown in
Figure \ref{fig:failure}(a),
for Shfflecast instances with $p=4$, $p=6$, and $p=8$, 75\%, 83\%, and 88\% of the source ToRs do not lose reachability from a relay failure, respectively. 
Moreover, we also observe that reachability is completely restored after enabling the single relay failure recovery (Algorithm \ref{alg:fail_recov}). Hence, Shufflecast is robust enough against single relay failure.

\textbf {b) Shufflecast has graceful performance degradation after failure recovery:} 
 According to lemma \ref{hop}, in a healthy $p,k$-Shufflecast any source ToR can reach all other ToRs within two complete traversals i.e., maximum hop count is $\left(2k-1\right)$. Based on our analysis, after enabling the failure recovery, the maximum hop count is unchanged for the majority of sources. Also, the upper bound of maximum hop count now becomes $\left(3k-1\right)$, i.e., any source ToR can reach all other ToRs within three complete traversals in the worst case. Figure \ref{fig:failure}(b)
demonstrates the CDF of possible increase in latency (in terms of maximum hop count) after single relay failure recovery for different Shufflecast instances ($p=4,6$ and $8$). 
We observe that, after single relay failure recovery the maximum hop count remains unchanged for $90-95\%$ of the sources and the possible increase in maximum hop count is upper bounded by $k=3$. 

In Figure \ref{fig:failure}(c) we vary the fraction of active ToR sources performing one-to-all multicast and observe the  multicast throughput degradation for different Shufflecast instances after enabling failure recovery. 
For a given fraction of active sources, we uniformly sample random set of ToRs and compute the relative multicast throughput degradation of those ToR sources between the healthy and failed network (after the failure recovery) averaged over the samples. The throughput for an individual ToR source is defined as the inverse of maximum fair-share for that source in presence of other active sources. As shown in 
Figure \ref{fig:failure}(c) shows that
 the average multicast throughput of Shufflecast degrades gracefully after failure recovery and the degradation reduces with bigger network scale. For Shufflecast instances with $p=4,6$ and $8$, the throughput degradation is upper-bounded by $25\%, 16.7\%$ and $12.5\%$ respectively. Such graceful degradation also reflects on the simultaneous multicast capability of Shufflecast. For a healthy $p,k$-Shufflecast, $p$ ToRs in one column, having their set of relays from disjoint partitions, can simultaneously perform a one-to-all multicast at line-rate. After the single relay failure, two partitions
of at least one column are shared, so the degree of parallelism now becomes $\left(p-1\right)$, i.e., $\left(p-1\right)$ ToRs can in parallel perform one-to-all multicast at line-rate.

\vspace{-3mm}
\subsection{Shufflecast achieves improved application performance for high-bandwidth and low-latency applications}
\label{high_thput}

\begin{figure*}[t!]
  \centering
    \includegraphics[width=6.9in]{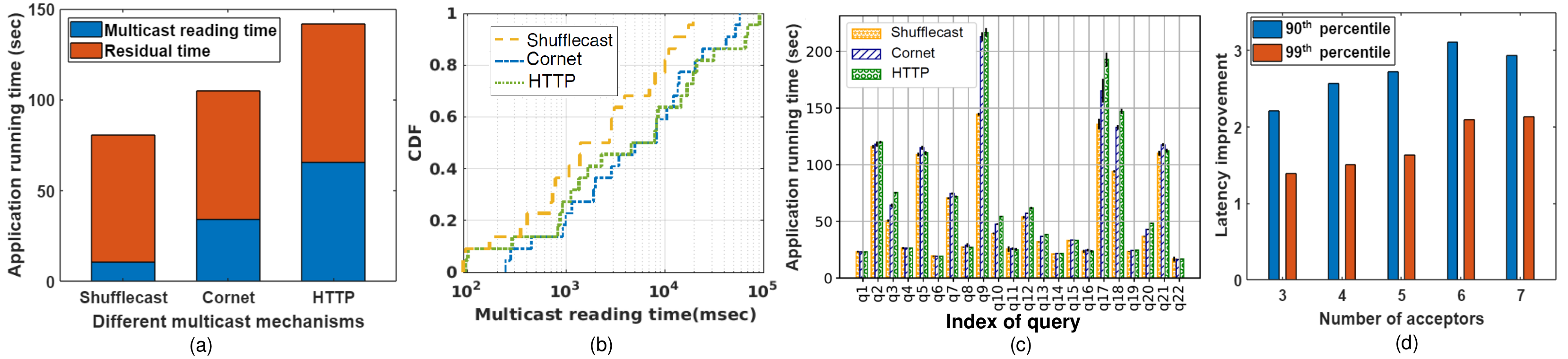}
    \caption{Application performance improvements of Shufflecast compared to native multicast mechanisms over full-bisection bandwidth network. (a) For LDA, the speedup in multicast reading time are $3.25\times$ and $6.24\times$ compared to Cornet and HTTP, respectively. The corresponding improvement in application running time are $23.41\%$ and $43.1\%$  (b) CDF of TPC-H multicast reading time. Shufflecast improves the distribution and achieves $2.7\times$ and $3.5\times$ speedup in total multicast reading time compared to Cornet and HTTP respectively, (c) The average running time of each TPC-H query (q$1$ to q$22$) over the three multicast mechanisms. For certain queries (e.g., $1$, $4$, $6$, $14$, $15$, $22$), the amount of multicast data is either very small (under $200$~MB) or non-existent, so there is no visible difference between Shufflecast, Cornet, and HTTP. However, for other queries (e.g., $9$, $17$, $18$), the multicast data is large ($5$~GB). Shufflecast improves the total query running time by $13.7$\% compared to Cornet and $17$\% compared to HTTP, (d) The latency improvements of multicast Paxos over Shufflecast compared to unicast Paxos over full-bisection bandwidth network with one sender and varying number of acceptors. The $90^{th}$ and $99^{th}$ percentile latency improvements are $2.21-2.93\times$ and $1.39-2.13\times$ respectively. }
    \label{fig:app_all}
\vspace{-2.5mm}
\end{figure*}

We briefly discuss three different workloads and experimentally demonstrate that real-world applications can leverage Shufflecast with only minor modifications.

\textbf{a) Spark ML:} Under Spark Machine Learning applications, we focus on Latent Dirichlet Allocation (LDA), one of the popular iterative machine learning algorithms. We use the Spark LDA implementation~\cite{cai2014comparison} with the dataset of $20$ Newsgroups as the input corpus~\cite{20newsgroups} which performs the one-to-all multicast for the training vocabulary model ($735$~MB in size). We use a cluster of $8$ servers to run LDA, where the application randomly chooses one server with four cores and $88$~GB RAM as the master, while the other seven servers with two cores and $44$~GB RAM serve as $14$ slave executors. Currently, the application uses Spark's native multicast mechanisms like Cornet~\cite{chowdhury2011managing} and HTTP (repeated unicasts to all receivers) over full-bisection bandwidth network. We use an extension to Spark that can perform multicast \cite{sun2018republic} over Shufflecast network and compare the application performance with Cornet and HTTP. We obtain the total multicast reading times and application running times averaged over $10$ runs, as shown in Figure~\ref{fig:app_all}(a). Shufflecast achieves $3.25\times$ and $6.24\times$ speedup in multicast reading time compared to Cornet and HTTP  respectively, with corresponding improvements of $23.41$\% and $43.1$\% in overall application runtime.

\textbf{b) Spark distributed database:} TPC-H is a widely used database benchmark of 22 business-oriented queries with high complexity and concurrent data modifications~\cite{TPCH}. We run these queries using the Spark SQL framework~\cite{chiba2016workload}. The database tables are $16$~GB in size overall, and the multicast data is one of such tables with size ranging from $4$~MB to $6.2$~GB for the distributed database join, making a total of $48.3$~GB of multicast data across queries. We compare the performance of TPC-H with and without Shufflecast keeping the same server configuration as Spark ML. Figure~\ref{fig:app_all}(b) shows the multicast reading time distribution of different multicast mechanisms across all TPC-H queries (queries $1$ to $22$). Shufflecast improves the distribution and gets speedup of $2.7\times$ and $3.5\times$ in total multicast reading time compared to Cornet and HTTP respectively. Figure~\ref{fig:app_all}(c) shows the application running time of each TPC-H query averaged over $10$ runs. For certain queries (e.g., $1$, $4$, $6$, $14$, $15$, $22$), the amount of multicast data is either very small (under $200$~MB) or non-existent, showing no visible difference between Shufflecast, Cornet, and HTTP. However, for other queries (e.g., $9$, $17$, $18$), multicast data is large ($5$~GB). The improvement of total query running time is $13.7$\% compared to Cornet and $17$\% compared to HTTP (Figure~\ref{fig:app_all}(c)).

\textbf{c) Paxos-based consensus protocol:} Paxos~\cite{lamport1998part, lamport2001paxos} is a consensus protocol that provides the foundation for building distributed fault-tolerant systems. Paxos has distributed entities called proposers, acceptors and learners. The execution of the protocol consists of four major steps, out of which three steps require one-to-many communications. As the messages tend to be small, the performance of Paxos is sensitive to latency. We run Paxos where the client repeatedly sends 1 Byte values to the proposer. The client sends the next value as soon as the previous is successful, and repeats for one hundred iterations; each iteration provides a latency measurement. All acceptors are placed on different servers. We run multicast-based Paxos \cite{LibFast} (natively leverage network-level multicast) over Shufflecast network (no application modification required) and compare the latency with unicast-based Paxos \cite{LibPaxos3} (repeated-unicasts to realize multicast) running over full-bisection bandwidth network, with one sender and varying number of acceptors.  Figure~\ref{fig:app_all}(d) shows that Shufflecast improves the tail latency significantly,  e.g., $90^{th}$ and $99^{th}$ percentile latency improvements are $2.21-2.93\times$ and $1.39-2.13\times$ respectively across different number of acceptors. 

\vspace{-6mm}
\section{Related Work}
\label{related}

Recent work has explored how software-defined networks (SDN) can be leveraged to improve IP-multicast support on packet-switched network 
(e.g. tree construction, group forwarding state maintenance, and
packet retransmissions) in the cloud data center setting, which is
related to the compute cluster
environment~
\cite{vigfusson2010dr,li2011scalable,li2012esm,li2013scaling,cao2013datacast,li2014reliable,shahbaz2019elmo}.
Our Shufflecast architecture directly connects the ToR switches which significantly reduces the excess resource usage. Also, shufflecast eliminates the need for run-time ToR-to-ToR-level multicast tree construction, group state exists
only at the network edge.
There have been proposals
\cite{WangH2012,CCR,samadi2014accelerating,blast,samadi2015optical} 
that use a MEMS-based OCS as a connectivity substrate to construct optical multicast trees 
via optical splitters. However, they are not scalable, they cannot achieve predictable performance, and they incur significant cost. Their
scalability is limited by the centralized OCS, which has only a few
hundred ports~\cite{Polatis384,Calient320,Glimmerglass192}, and these
ports need to interconnect all ToRs and all in/out ports of optical
splitters. Scalability is further limited by the need for optical
power amplification, which is difficult and expensive when the tree
gets large. The performance predictability of these proposals is hurt
by long circuit switch configuration delays that are exacerbated by
the need to concatenate multiple optical circuits through splitters to
form the tree. Moreover, OCS incurs significant cost which restricts such proposals from large scale deployment. 
In contrast, Shufflecast provides simple, 
scalable and data-rate agnostic multicast in a more power efficient and economical way.
\cite{wu2016hyperoptics} proposes a topology
that eliminates the centralized OCS, but its scalability is inherently
limited by splitter fan-out and the entire proposal consists of only
the topology design. In contrast, Shufflecast's topology can scale to
an arbitrary size even with a small splitter fanout and we have
demonstrated the complete system's effectiveness using end-to-end
applications.

\vspace{-4mm}
\section{Conclusion}
\label{conclusion}

Optical circuit-switched (OCS) network core has several advantages to be a potential candidate for next-generation compute clusters. However, there is no inherent support for multicast by such networks. Our proposed Shufflecast architecture can complement those high performance OCS-based core and support data-rate agnostic multicast maintaining low power and low capital cost. Shufflecast's data plane is scalable and supports line-rate throughput; its control plane is simple and responsive; Shufflecast is robust enough against failure. Experiments using a complete hardware and software prototype of Shufflecast show that Shufflecast can improve the performance of real-world applications with minor modifications.

\bibliographystyle{abbrv}
\begin{small}
\bibliography{reference}
\end{small}
\vspace{-2mm}
\appendices
\section{}
\label{appendix}

\vspace{-2mm}
\noindent
\subsection{Details of Shufflecast Data Plane}
\label{hopsProof}
\subsubsection{Scalability of Shufflecast fabric} 
\label{$(2,3)$-Shuffle}
Shufflecast can scale easily even with small fanout splitters.
Figure~\ref{fig:connectivity_big} shows  $2,3$-Shufflecast consisting of $3\cdot2^3=24$ ToRs arranged in $3$ columns connected via $1:2$ optical splitters, and each column has $2^3=8$ ToRs. We can realize even bigger instances of Shufflecast with small $p$. For example, $4,4$-Shufflecast uses $1:4$ splitters, covering $1024$ ToRs. 

\vspace{-2mm}
\subsubsection{Detailed Analysis of Multicast-aware routing} 
\label{Routing_detail}
First, the next-hop relay computation algorithm (Algorithm \ref{alg:bcast_route}) computes the column-difference parameter $X \left(<=k\right)$ between the destination ToR ($dest$) and the current ToR ($cur$: initialized to $src$). If both the ToRs belong to the same column, $X$ is considered as $k$ (lines $3$ and $4$), otherwise $X\left(<k\right)$ is computed as stated by line $6$. From the construction of Lemma \ref{hop}, we observe that $X$ dictates the hop count from $cur$ to $dest$.
If both the ToRs belong to the same column, both $X$ and the hop count are $k$ (reachable at the end of the first cycle). Otherwise, 
$dest$
is reachable either in hop count $X$ (during the first cycle) or $k+X$ (during the second cycle). Next, the algorithm checks whether the hop count from 
$cur$ 
to $dest$ 
is $<=k$ (line \ref{cond0}) by matching their partial row-ID digits ($k-X$ most and least significant row-ID digits of the 
$dest$ and $cur$ 
 respectively). Finally, the next-hop ($next$) ToR ID is determined by shifting the current ToR's row-ID to the left by one digit, and then putting the $\left(X-1\right)^{th}$ digit of the destination ToR's row-ID (line \ref{op0}) if the condition is true, or putting the $\left(k-X'-1\right)^{th}$ digit of the source ToR's row-ID (line \ref{op1}) if the condition is false, where $X'\left(<k\right)$ is the column-difference parameter between 
$cur$ and $src$ (line $11$).

\vspace{-2mm}

\subsubsection{Proofs of Lemmas} 
\label{lemma_proof_appen}
\textbf{Proof of Lemma \ref{hop}:}
By construction of a $p,k$-\sys, any given source ToR has $1:p$ splitter connecting $p$ ToRs of the next column in \nth{1} hop, again from those $p$ ToRs another $p^2$ ToRs at two-columns ahead from the source are reachable in \nth{2} hop and so on. Eventually $p^{k-1}$ ToRs belonging to one partition at previous column of source are reachable in $\left(k-1\right)$ hops which is sufficient for reaching all $p^k$ ToRs of its own column in the next $k^\text{th}$ hop. 
During the second  cycle, the remaining ToRs of next column from the source are all reachable from any of the partitions of $p^{k-1}$ ToRs at source column. The same scenario follows for all the consecutive columns during the second cycle, reaching the remaining ToRs of all the other columns. Finally, the remaining ToRs at the previous column of source can be reached in another $\left(k-1\right)$ hops. Therefore, all the ToRs are reachable within two cycles of traversal i.e., the hop count is at most $\left(k+k-1\right)=2k-1$.

\textbf{Proof of Lemma \ref{p_criteria}:} 
By construction of $p,k$-Shufflecast, for the destinations reachable in at most $k$ hops (i.e., during the first cycle), the chosen relays are at most $\left(k-1\right)$ hops away from the source, with most significant digit as source row-ID digits left shifted by at most $\left(k-1\right)$ places. 
 As a result, the relays are inherently chosen from the partition IDs defined by the source row-ID digits. Hence, appending the pre-calculated destination digit ($r^d_{X-1}$) as the least significant digit ensures the shortest-path next-hop relay ID following the \textit{partition criteria}. After first cycle, all the $k$ source row-ID digits are ignored due to $k$ effective left shifts. Therefore, for all the remaining ToRs reachable in the second cycle, the algorithm ensures the \textit{partition criteria} by appending the pre-calculated source row-ID digit ($r^s_{k-X'-1}$) as the least significant digit during the first cycle. These digits govern the selective choice of relays from proper partition IDs during the second cycle. Hence, any given source ToR can perform one-to-all multicast following the partition criteria. 

\textbf{Proof of Lemma \ref{simul_bcast}:}
Following the partition criteria in \ref{p_criteria}, a given source ToR ($c^s,r^s_{k-1},r^s_{k-2}...r^s_{1}r^s_{0}$) in a $p,k$-\sys performs one-to-all multicast using relays from its own column with partition ID $r^s_{k-1}$, from next column with partition ID $r^s_{k-2}$ and so on, finally from previous column with partition ID $r^s_0$. We also know, each column contains $p$ partitions as every row-ID digit can have $p$ distinct values ($\in\left[0,p-1\right]$). Eventually, to perform one-to-all multicast at line-rate, the group of source ToRs are to be chosen so that the relays are disjoint i.e., from distinct partitions at every column. Thus for the given source, the group of other source ToRs from the same column must have all distinct $k$ row-ID digits. 
Intuitively, we must choose one ToR from each of the $p$ partitions which at least makes all the most significant digits distinct. 
For example, given source ToR row-ID, if we choose one $j \in \left[0,p-1\right]$ and perform $\left(r^s_i + j\right) \bmod p$ for all $i \in \left[0,k-1\right]$, eventually we get $p$ ToRs having all distinct $k$ row-ID digits and hence they can perform one-to-all multicast simultaneously at line-rate using relays from distinct partition. Now, if we choose two such groups of $p$ ToRs, effectively we have two ToRs from each of the $p$ partitions. Thus, for each of the $k$ places, there exist two unique ToRs using the same digit twice a given place which results them uniquely sharing the relays from same partition. Hence, those $2p$ ToRs can make one-to-all multicast simultaneously at half of the line-rate. Extending this idea, we can choose all such $p^{k-1}$ groups of $p$ ToRs i.e., all the $p^k$ ToRs of one column using the relays from same partition and hence they can make one-to-all multicast at $p^{k-1}$ fraction of the line-rate.

\textbf{Proof of Lemma \ref{optimality}:}
In a $p,k$-\sys, every ToR of a given column is connected to another $p$ ToRs of its next column, and every column has $p^k$ ToRs. Therefore, we need at least $p^{k-1}$ ToRs of a given column to reach all the ToRs of the next column. Hence a given source ToR must require at least $p^{k-1}$ number of relays from each of the $k$ column to perform one-to-all multicast. In Lemma \ref{p_criteria} we have already proved, with multicast-aware routing any source ToR can perform one-to-all multicast using the relays from one partition at each column. From the definition of partition we know, every partition has $p^{k-1}$  ToRs which is the same as the minimum relay requirement. Thus, multicast-aware routing minimizes the relay usage. Also, we know there are $p$ partitions per column. Hence, with such minimum relay requirement, maximum $p$ sources in one column can possibly use disjoint set of relays from every column and consequently can perform one-to-all multicast simultaneously at line-rate. This is indeed the number of simultaneous one-to-all multicast supported by multicast-aware routing at line-rate as proved in Lemma \ref{simul_bcast}. Thus, multicast-aware routing is optimal in terms of relay usage and multicast performance. 

\noindent
\subsection{Analysis to show unused ToR ports often exists}
\label{appendixunusedport}
\begin{figure}
  \centering
  \setlength\belowcaptionskip{-1\baselineskip}
\begin{tikzpicture}
  \node (img1) {\includegraphics[width=2.5in, height=1.8in]{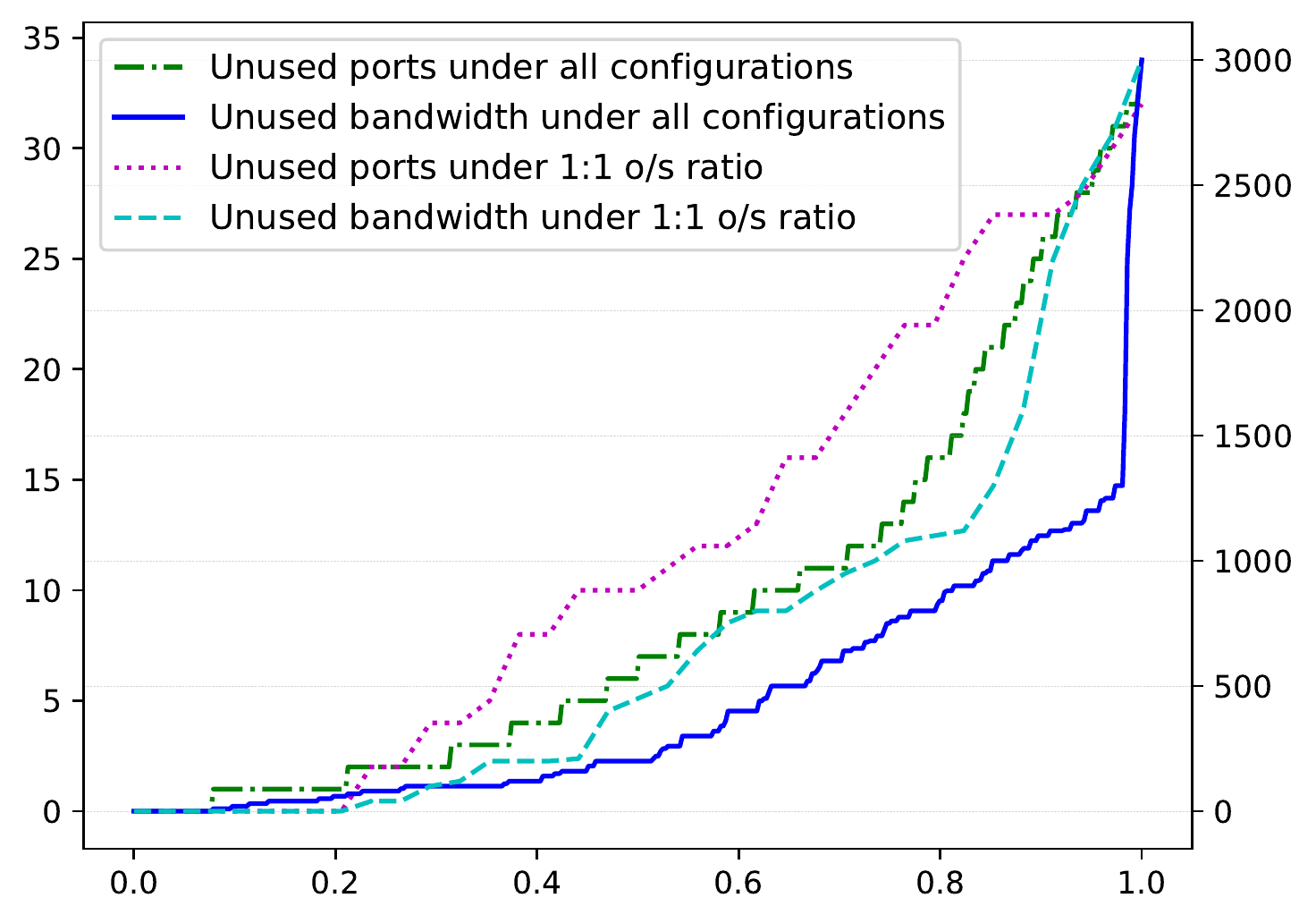}};
  \node[below=of img1, node distance=0cm, yshift=1.2cm] {CDF};
  \node[left=of img1, node distance=0cm, rotate=90, anchor=center,yshift=-0.9cm] {\small The number of unused ports};
  \node[right=of img1, node distance=0cm, rotate=90, anchor=center,yshift=0.9cm] {\small The unused bandwidth (Gbps)};
\end{tikzpicture}
\vspace{-2mm}
\caption{CDF of unused ports and unused bandwidth under all configurations and $1:1$ oversubscription ratio configuration.}
\label{fig:unused ports CDF}
\vspace{-2mm}
\end{figure}
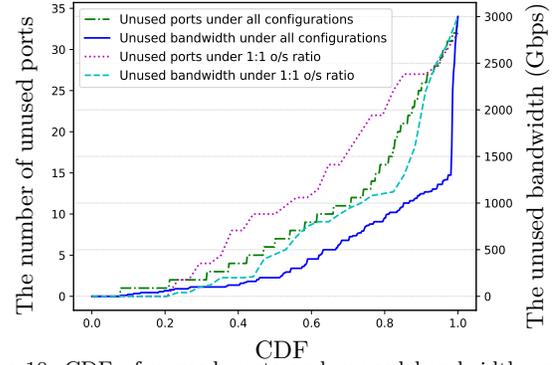

Our methodology considers a wide range of network configurations. For each configuration, we choose the ToR switch that minimizes the amount of unused bandwidth. 
We study $14$ types of ToR switches with different port configurations
from several well-known companies. 
Specifically, we use $2$ HP switches with either $24\times10$
Gbps ports or $48\times10+4\times40$ Gbps ports, $2$ Juniper switches
with either $32\times10$ Gbps ports or $48\times10+4\times100$ Gbps
ports, $3$ Arista switches ranging from $32\times10+4\times40$ Gbps
ports to $96\times10+8\times40$ Gbps ports, and $8$ Cisco switches
ranging from $32\times40$ Gbps ports to $64\times100$ Gbps ports.
For the network configurations, we adopt several oversubscription
ratios reported in the literature, i.e., $1:1$, $3:2$, $3:1$, $4:1$,
$5:1$, $8:1$, $10:1$ and
$20:1$~\cite{wu2019say,greenberg2009vl,impactPS,Cao:15}. We also
include a few additional oversubscription ratios: $x:1$ where $x \in [1,10]$. 
We consider commercially available standard rack cabinet sizes ranging from $18$U to $48$U~\cite{Racksolutions,HPE}, and five different per-server network port speed configurations -- $10$ Gbps, $2\times10$ Gbps, $25$ Gbps, $40$ Gbps and $2\times25$ Gbps.
The detailed results are shown in Figure~\ref{fig:unused ports CDF}. Indeed,
unused ports, as well as a large amount of unused bandwidth, often
exist. Among all cases, the configuration of $1:1$ oversubscription is unique and the unused ToR ports truly cannot be used to add more
bandwidth into the network core.

\vspace{-2mm}
\setlength{\belowcaptionskip}{-10pt}

\end{document}